\documentclass[amsmath,amssymb,prd,superscriptaddress,a4paper,nofootinbib,11pt]{article}\pdfoutput=1
\usepackage{jheppub}
\usepackage{amsthm}
\usepackage{float}
\usepackage{graphicx}
\usepackage{epstopdf}
\usepackage{color}
\usepackage{multirow}

\newcommand{\be}{\begin{equation}}
\newcommand{\ee}{\end{equation}}

\def\bsp#1\esp{\begin{split}#1\end{split}}
\def\bpm{\begin{pmatrix}}
\def\epm{\end{pmatrix}}

\newcommand{\bea}{\begin{eqnarray}}  
\newcommand{\eea}{\end{eqnarray}}  

\def\lsim{\raise0.3ex\hbox{$\;<$\kern-0.75em\raise-1.1ex\hbox{$\sim\;$}}}
\def\gsim{\raise0.3ex\hbox{$\;>$\kern-0.75em\raise-1.1ex\hbox{$\sim\;$}}}

\title{Towards model-independent exclusion of light Stops}
\author[a,b,1]{Alexander Belyaev\note{e-mail: a.belyaev@soton.ac.uk}}
\author[c,2]{Ver\'onica Sanz\note{e-mail: v.sanz@sussex.ac.uk}}
\author[a,b,3]{Marc Thomas\note{e-mail: m.c.thomas@soton.ac.uk}}

\affiliation[a]{School of Physics and Astronomy, University of Southampton, Highfield, Southampton SO17 1BJ, UK}
\affiliation[b]{Particle Physics Department, Rutherford Appleton Laboratory, Chilton, Didcot, Oxon OX11 0QX, UK}
\affiliation[c]{Department of Physics and Astronomy, University of Sussex, Brighton BN1 9QH, UK}

\abstract{Understanding the extent to which experimental searches are sensitive to Light Stops (LST) scenarios is essential to resolve questions about naturalness, electroweak baryogenesis and Dark Matter. In this paper we characterize the reach on LST scenarios in two ways. We extend experimental searches to cover specific gaps in the LST parameter space, showing for the first time that assuming a single decay channel one can exclude the region of $m_{\tilde{t}}<m_{top}$, which in its turn excludes electroweak baryogenesis in MSSM.
Also, we explore the extent to which searches are weakened in a more generic scenario when more than one decay channel takes place, even after their combination. This study highlights the need for a more comprehensive exploration of the LST parameter space.}

\begin{document}
\maketitle
\flushbottom
\newpage

\section{Introduction\label{sec:Intro}}

Supersymmetry (SUSY)~\cite{Golfand:1971iw,Ramond:1971gb,Neveu:1971rx,Gervais:1971ji,Volkov:1973ix,Wess:1973kz} is one of the most favoured theories beyond the Standard Model (SM).  It naturally solves the hierarchy problem, provides  gauge coupling unification and can address fundamental experimental problems of the SM on the cosmological scale, such as Dark Matter(DM) and Electroweak Baryogenesis. SUSY enlarges the SM spectrum of particles by their superpartners (sparticles), in particular with scalar partners of SM fermions -- sfermions. There are two scalar partners for each SM fermion, one corresponding to each chirality, conserving the number of degrees of freedom. Among them, stop quarks(stops), $\tilde{t}_{1,2}$, the super partners of the top-quark play a special role.

First of all $\tilde{t}_{1,2}$ controls the radiative corrections to the Higgs boson mass, measured to be around 125~GeV~\cite{CMS:2014ega,Aad:2014aba}. For the minimal supersymmetric standard model (MSSM) this implies  that the mass of at least one of  $\tilde{t}_{1,2}$ should be {\it of the order of above} a TeV scale, since radiative corrections are proportional to the  $\log$ of $\tilde{t}_{1,2}$, and are required to be large because the tree-level MSSM mass is below $M_Z$. 
This puts the mass of $\tilde{t}_{1,2}$ in tension with SUSY naturalness which suggests 
that the third generation squarks should be {\it below about 1 TeV}~\cite{Ellis:1986yg,Barbieri:1987fn}. This follows from the simple equation
\begin{equation}
\frac{M_Z^2}{2} \simeq M_{H_u}^2-\mu^2
\label{eq:ft}
\end{equation}
which connects the $Z$-boson mass, $M_Z$, the radiatively corrected $H_u$ mass term of the 
superpotential, $M_{H_u}$ (which depends on the stop contributions) and the superpotential higgsino mass, $\mu$.
It has however been shown that the usual fine-tuning measures, defined as the sensitivity of the weak scale to fractional variations in the fundamental parameters of the theory, can be low even if the masses of the supersymmetric scalars 
are large. 
This happens in the so called ``hyperbolic branch"(HB)~\cite{Chan:1997bi} or ``focus point" 
(FP)~\cite{Feng:1999mn,Feng:1999zg,Feng:2011aa} regions of the minimal super gravity (mSUGRA) parameter space, where the value of the Higgs mass parameter, $\mu$, can be low if the universal gaugino mass $M_{1/2}$ is not too large.
Moreover it was recently argued~\cite{Baer:2013gva} that EW fine-tuning in SUSY scenarios 
can be grossly overestimated by neglecting additional terms, stemming from the ultra violet (UV) completion of the model, that can lead
to large cancellations favouring a low $\mu$-parameter and allowing a heavier stop mass (up to a certain limit).

Besides its connection to the Higgs boson mass and fine-tuning,
the light stops can also affect the Higgs signal at the LHC, namely by altering its production via gluon-gluon fusion
and decay branching ratios, which was the subject of many detailed studies, see e.g.~\cite{Espinosa:2012in,Altmannshofer:2012ks,Carena:2013iba,Belyaev:2013rza,Fan:2014txa,Katz:2015uja} for light stops, Refs.~\cite{Henning:2014wua,Drozd:2015kva,Huo:2015nka} for studies in the context of Effective Field Theories, and \cite{Belanger:2015vwa} for a study where flavour and Dark Matter constraints are also considered.
In particular, it was shown~\cite{Belyaev:2013rza} that a scenario with light stops, would be able 
to explain a non-universal alteration of the two most relevant Higgs production channels ---
gluon-gluon fusion and vector boson fusion ones.

Finally, one should note that the light
stop scenario  is also attractive from a cosmological point of view. Firstly, there is a scenario where the lightest neutralino, being Dark Matter (DM) is degenerate with the lightest stop in the 100-300 GeV range, which predicts a plausibly low amount of DM (via the stop-neutralino co-annihilation channel)~\cite{Boehm:1999bj,Ellis:2001nx,Balazs:2004ae,Ellis:2014ipa,deSimone:2014pda}, and secondly the light stop scenario enables Electro-Weak Baryo-Genesis (EWBG) by facilitating a first-order phase transition, specifically requiring very
light stops ($m_{\tilde{t}}<150$~GeV)\cite{Carena:1997gx}.
It should be noted that a number of papers claim to have ruled out light stop mediated EWBG in the MSSM by setting limits on stop masses using Higgs data \cite{Katz:2015uja,Cohen:2012zza,Curtin:2012aa}. However others have found loop holes in such arguments, such as the case where neutralinos have masses below about 60 GeV causing a sizeable Higgs decay to invisible \cite{Carena:2012np} or in the so-called {\it funnel region}~\cite{Espinosa:2012in} where the two stops could conspire to eliminate their effect on Higgs couplings. 
One should also note that the LST exclusion of 100 GeV stops using  the Higgs data
is substantially  weakened in case of  $m_{\tilde{t}}\simeq 150$~GeV
which still could possibly trigger EWBG.
Therefore ruling out of light stops via direct searches is an important step in 
straigthforward exclusion of  EWBG, as this way  is independent of Higgs measurements and closes  loopholes mentioned above.

\begin{figure}[htb]
\begin{center}
\includegraphics[width=0.62\textwidth]{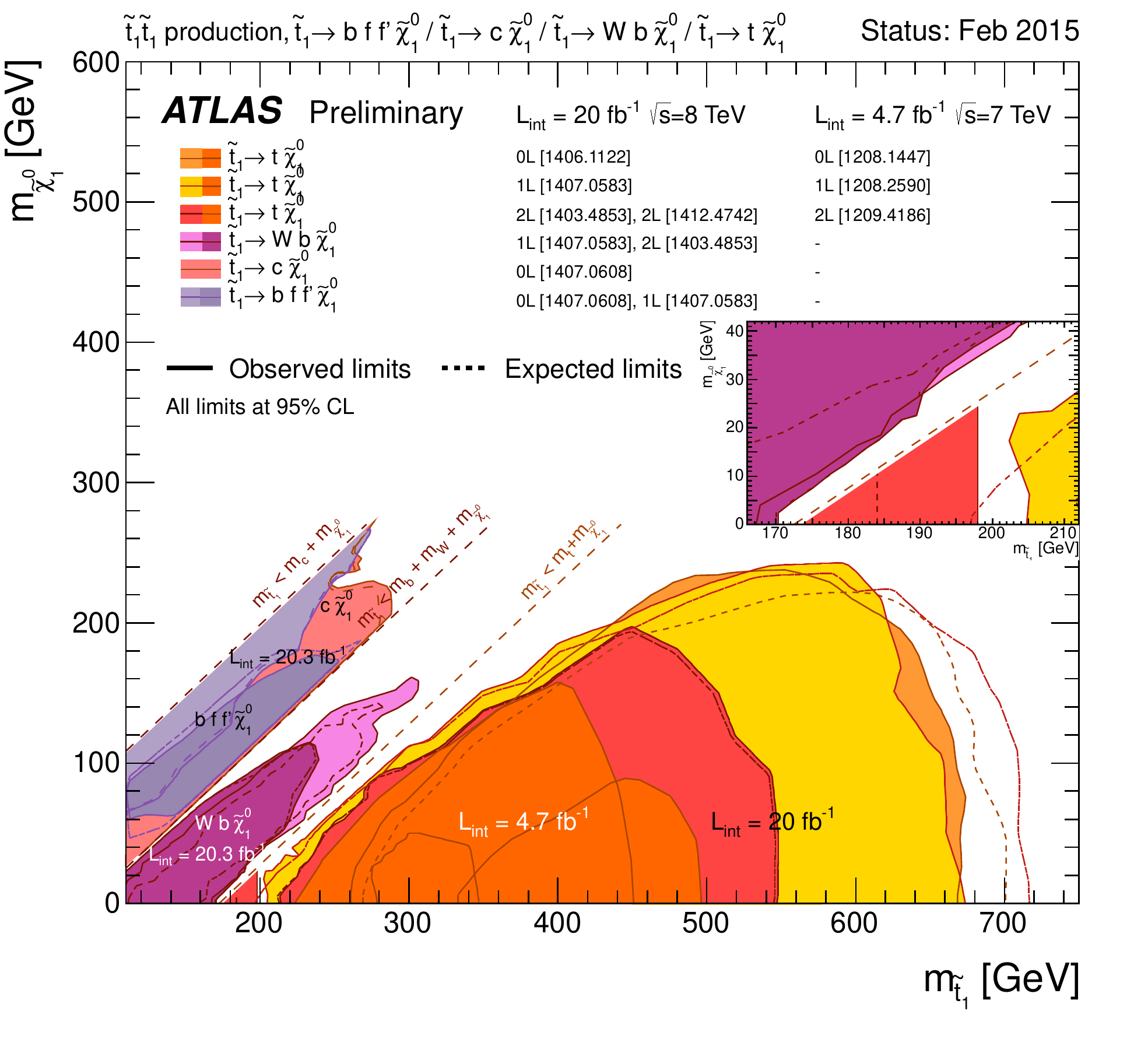}\\
\includegraphics[width=0.62\textwidth]{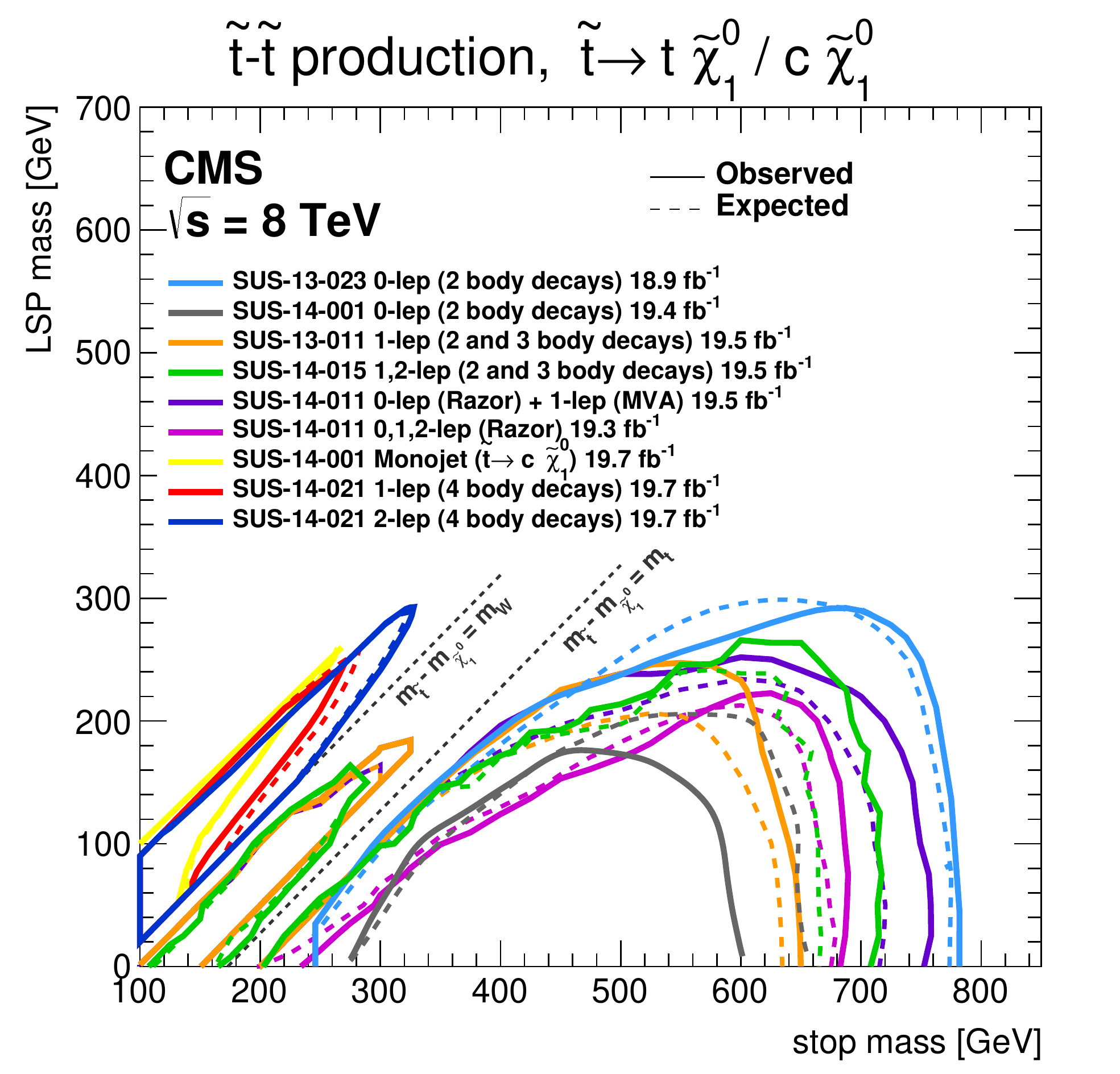}
\caption{Combined results on stop searches by ATLAS(top)~\cite{atlas-stop} and CM(bottom)~\cite{cms-stop} collaborations:
presenting the 95\% confidence limits (CL) exclusion region in the stop mass ($m_{\tilde{t}}$) vs neutralino mass ($m_{\tilde{\chi}^0_1}$) plane. Each search assumes a 100\% branching ratio via certain channels as shown in the legend of the plot
\cite{ATLAS_Twiki,Aad:2012ywa,Aad:2012xqa,Aad:2012uu,Aad:2014kra,Aad:2014bva,Aad:2014qaa,Aad:2014mfk,Aad:2014nra}.
\label{stop-atlas-cms}}
\end{center}
\end{figure}

One can see that the answer to the question of what is the lower stop mass limit is  
crucial for  different aspects of Supersymmetry -- one of the most appealing BSM theories. 
Currently the  light stops mass range below 300 GeV is  highly restricted by present experimental data, but 
not fully excluded. For example, the region near the thresholds of stop decay 
to  $Wb\tilde{\chi}^0$ or to  $t\tilde{\chi}^0$  has not been fully covered neither by ATLAS nor by  CMS collaborations.
This can be seen  from  ATLAS and CMS  combined limits on the stop mass in the neutralino-stop mass plane presented in Fig.~\ref{stop-atlas-cms}  which
is taken from  Refs.~\cite{atlas-stop,cms-stop}.
From this figure one can see that indeed for the $\Delta m = m_{\tilde{t}}-m_{\tilde{\chi}^0_1} \gtrsim m_b+m_W$,
i.e. just below the line indicating the $Wb\tilde{\chi}^0$ threshold, there is a small sleeve of the allowed parameter space 
for  $m_{\tilde{t}}$ even below 150~GeV. 
We should note that while finalising this paper we were made aware of a phenomenological paper \cite{Rolbiecki:2015lsa} which improves on the limits in this region using precision measurements of the $W^+ W^-$ cross section. Our approach is different and complimentary, and also more general as we also explicitly consider large branching ratios to the 2-body $\tilde{t}_1 \to c\tilde{\chi}^0_1$ final state.

When $\Delta m <  m_b+m_W$, the stop can decay via a
radiatively induced $\tilde{t}_1 \to c\tilde{\chi}^0_1$ 2-body (2BD) flavour violating channel which can be large or even dominant in this  $\Delta m$ region. Moreover,  in the very narrow region
$m_c<\Delta m <  m_b$ (not indicated in Fig.~\ref{stop-atlas-cms}) this is the only possible decay channel. The other possible decay in the  $\Delta m <  m_b+m_W$
parameter space is the 4-body stop decay channel (4BD) $\tilde{t}_1 \to b f \bar{f'} \tilde{\chi}^0_1$,
were $f$ and $f'$ are either quarks of the 1st or 2nd generation, or leptons.
This decay is realised either via the exchange of a virtual top quark and W-boson and/or a virtual chargino.
In this paper we are interested in the LHC sensitivity to the light stop (LST)
scenario, which  we define hereafter as $m_{\tilde{t}}<m_{top}$ to be concrete.
One can see in Fig.~\ref{stop-atlas-cms} that the LST scenario can be 
excluded assuming a 100\% branching ratio to 2BD decays using monojet and 
charm-tagged monojet signals~\cite{Aad:2014nra} or under the assumption of a 100\% branching ratio to 4BD decay 
using a combination of monojet-like signal selection~\cite{Aad:2014nra} (which is also studied by CMS~\cite{CMS:2014yma}) and  single lepton plus missing transverse momentum signatures~\cite{Aad:2014kra}.
The very important question is the status of  the LST scenario
in the case of mixed branching ratios. This can be parameterised by a single parameter 
\begin{equation}
\epsilon_{2B}=Br(\tilde{t}_1 \to c\tilde{\chi}^0_1)
\end{equation}
so,
\begin{equation}
Br(\tilde{t}_1 \to b f \bar{f'} \tilde{\chi}^0_1) =1-\epsilon_{2B}.
\end{equation}
This is one of the objectives we study in our paper.
Indeed, this simple parameterisation is possible in the $\Delta m <  m_b+m_W$
region since no new kinematics is expected  in the form of on-shell lightest chargino decay since its mass is limited by LEPII 
to be above about 88 GeV\cite{Heister:2002mn,Barate:1997dr,Heister:2002vh,Barate:1999fs}, so $\Delta m <  m_b+m_W < m_{\tilde{\chi}^\pm_1}$.

\begin{figure}[htb]
\begin{center}
\includegraphics[width=0.8\textwidth]{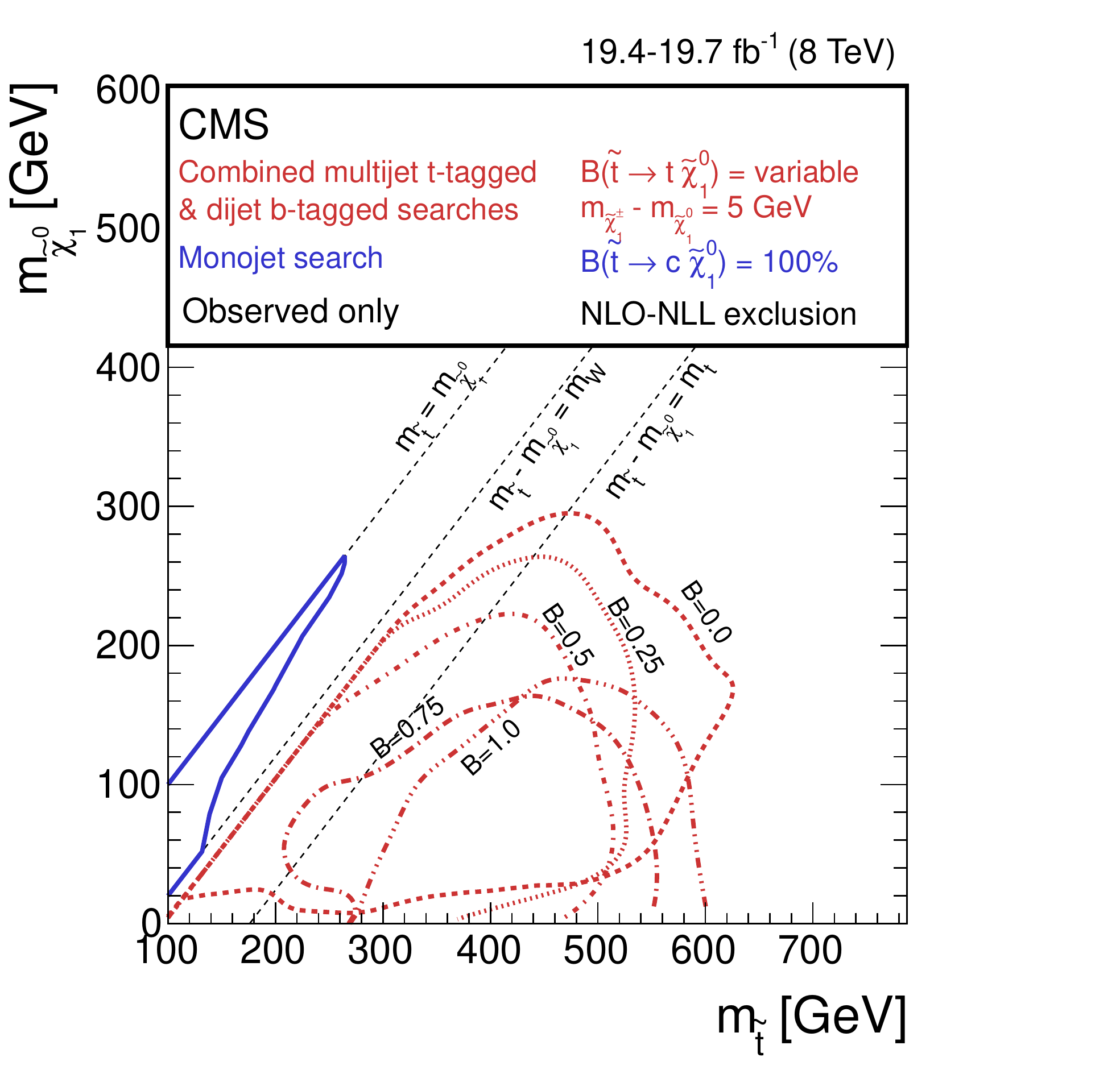}\\
\caption{Observed 95\% CL mass exclusion limit curves for top-squark pair production,
taken from Fig.11 of CMS SUS-14-001 paper\cite{Khachatryan:2015wza}
\label{stop-cms-alljets}}
\end{center}
\end{figure}

In the  $ m_b+m_W < \Delta m <  m_t$  region
the model-independent exploration of LST scenario is much more complicated as it can
involve the new kinematics from  the on-shell chargino decay.
In general, one needs to involve two  more parameters in this region -- the chargino mass and 
the branching ratio of decay via the on-shell chargino,
$ Br(\tilde{t}_1 \to  b \tilde{\chi}^{\pm}_1 \to b W^{\ast} \tilde{\chi}^0_1)$.
This mass region was explored under different assumptions by ATLAS
using 1- and 2-lepton signatures\cite{Aad:2014kra,Aad:2014qaa} and CMS using a 1-lepton signature\cite{Chatrchyan:2013xna}
and a signature involving fully hadronic final states from stops decay~\cite{Khachatryan:2015wza}.
The CMS collaboration has parameterised chargino mass using a parameter $x$ $(0,1.)$
\begin{equation}
m_{\tilde{\chi}^\pm_1}=x m_{\tilde t} + (1-x) m_{\tilde{\chi}^0_1}
\end{equation}
which defines the ``position'' of the chargino  between the stop and the lightest neutralino.
ATLAS and CMS studies in the $ m_b+m_W < \Delta m <  m_t$  region involving different signatures are very complementary
and very comprehensive,
but unfortunately they do not give a definitive answer as to whether the LST scenario is excluded or not over the whole MSSM parameter space. For example, the CMS study of fully hadronic final states from stops decay~\cite{Khachatryan:2015wza}
clearly demonstrates the importance of this signature to exclude the LST scenario
as indicated in Fig.~\ref{stop-cms-alljets}.
However, it's important to note that the results of this study are very model dependent:
when $Br(\tilde{t}\to b\tilde{\chi}^\pm_1) \equiv 1-Br(\tilde{t}\to t\tilde{\chi}^0_1) = 100\%$,
this search maximally excludes the $ m_b+m_W < \Delta m <  m_t$ 
region, while in the opposite case when $Br(\tilde{t}\to t\tilde{\chi}^0_1) = 100\%$ it leaves this
region intact. 

This happens because this study only considers 
on-shell top-quarks for $\tilde{t}\to t\tilde{\chi}^0_1$ decays which do not occur when $ m_b+m_W < \Delta m <  m_t$,
contrary to the $\tilde{t}\to b\tilde{\chi}^\pm_1$ decays for which the 
$\tilde{\chi}^\pm_1$ is always on-shell
since the mass gap between the chargino and neutralino was fixed at 5 GeV.

One can see that in order to perform a generic exploration
of the LST scenario one should complete the study of the respective space. As discussed above, this parameter space is three-dimensional  in $\Delta m <  m_b+m_W$ region
($m_{\tilde{t}}, m_{\tilde{\chi}^0_1}$ and $\epsilon_{2B}$),  
and five-dimensional in the  $ m_b+m_W < \Delta m <  m_t$ region
($m_{\tilde{t}}, m_{\tilde{\chi}^0_1}, \epsilon_{2B}, m_{\tilde{\chi}^\pm_1}$ and $ Br(\tilde{t}_1 \to  b \tilde{\chi}^{\pm}_1 \to b W^{\ast} \tilde{\chi}^0_1)$).
One can see that a model-independent exploration of the LST scenario is quite a complicated task.
On the way towards completing this task we make a simplification and consider only the part of the LST parameter space 
assuming that the chargino is heavier than the stop: $ m_{\tilde{\chi}^\pm_1}>m_{\tilde{t}}$.
This assumption allows us to conveniently work in a three-dimensional space in both the
$\Delta m <  m_b+m_W$ and  $ m_b+m_W < \Delta m <  m_t$ cases i.e. in the whole LST parameter space.
Relaxing this assumption would be the topic of future work, and in what follows we assume $ m_{\tilde{\chi}^\pm_1}>m_{\tilde{t}}$.

In this paper we explore the full LST
parameter space ($m_{\tilde{t}}, m_{\tilde{\chi}^0_1}$ and $\epsilon_{2B}$)
(under the $m_{\tilde{\chi}^\pm_1}>m_{\tilde{t}}$ assumption) using simulations with stops decaying via an off-shell top-quark and a W-boson
and recasting existing ATLAS and CMS searches. This allows us to explore the RUN1 LHC sensitivity to the full LST parameter space which is the main new  result of this paper.
It is clear that the potential exclusion of the whole LST parameter space
would have a dramatic consequences for the  EWBG scenario and important 
connections to Higgs phenomenology.

%\AB{Todo:
%\\
%1. review Maggie papers
%\\
%2. Write about the improvement we did
%-- the exact ME simulation
%to cover difficult region, which makes the link to model-independent 
%coverage of the  $\Delta M$ parameter space.
%\\
%3. that we reproduce Exp results and go further establishing ours;
%that we made an attemt to understand the reson of the existing gaps and 
%aimed to improve the present experimental interpretation,
%making path towards mode-independent coverage $\Delta M$ parameter space
%Notes:
%ATLAS: 
%}

The rest of the paper is organised as follows.
In Section II we discuss the stop decay channels and their dependence on the MSSM parameters.
We continue in Section III to discuss experimental analyses for the 2BD $\tilde{t}_1 \to c\tilde{\chi}^0_1$,
and 4BD $\tilde{t}_1 \to b f \bar{f'} \tilde{\chi}^0_1$ channels, and the tools and framework with which 
we extend these.
In section IV we present our results on the extension of the current experimental analyses and demonstrate that
for a generic case with intermediate branching ratios to the 2BD and 4BD decays, the current stop mass
limits are drastically reduced.
We draw our conclusions it Section V.

\section{Stop decay channels\label{sec:StopDecays}}

As discussed in the introduction, the LST scenario we study here
($ m_{\tilde{\chi}^\pm_1}>m_{\tilde{t}}$) is described by a three-dimensional 
($m_{\tilde{t}}, m_{\tilde{\chi}^0_1}$ and $\epsilon_{2B}$) parameter space.
One should note that for $\Delta m > m_b+m_W$ the stop will undergo a 3-body decay via a real W-boson.
However hereafter we will call any stop that decays via $\tilde{t}_1 \to b W^{(*)} \tilde{\chi}^0_1 \to b f\bar{f'} \tilde{\chi}^0_1$ a 4-body (4BD) channel since it leads
to a 4-body final state, originating either from a real ($\Delta m > m_b+m_W$) or virtual  ($m_b < \Delta m < m_b+m_W$)
W-boson.

The take-home message of this section is that neither 2BD nor 4BD channels are necessarily dominant in a given region of the parameter space. Therefore, it is crucial to understand how LHC limits change when these two channels compete  with each other.  
The role of these channels was a subject of several  studies since about decade from 
now~\cite{Hiller:2008wp,Hiller:2009ii,Muhlleitner:2011ww,Krizka:2012ah,Belanger:2013oka,Grober:2014aha,Grober:2015fia}, and in the paper we discuss how these studies compare with the present one.

Let us start the discussion with 2BD, which come from the flavour non-diagonal interaction of the stop with the charm and neutralino. The general form of $\tilde{t}_1-c-\tilde{\chi}^0_1$ interactions is~\cite{Hiller:2008wp}
$$\mathcal{L}_{\tilde{t}_1 c \tilde{\chi}^0_1} = \bar{c} (y_L P_L +y_R P_R)\tilde\chi^0_1 \tilde{t}_1  +h.c.$$
and the respective $\tilde{t}_1 \to c\tilde{\chi}^0_1$ decay width is given by
$$\Gamma_{2BD} =\frac{Y^2}{16\pi} \left( 1-\frac{m_{\tilde{t}_1}}{m_{\tilde{\chi}^0_1}} \right)^2 \,  m_{\tilde{t}}$$
with $Y=\sqrt{|y_L|^2+|y_R|^2}$. Approximate expressions for the couplings $y_{L,R}$ can be obtained by expanding in the charm mass~\cite{Krizka:2012ah,Grober:2014aha,Grober:2015fia} leading to
\bea
y_L = c_{SUSY} \,  \frac{y_b^2 V_{cb} V^*_{tb}}{m_{\tilde t}^2-m_{\tilde c}^2} \, \left(g' N_{11}+ 3 g N_{21} \right)\label{yL}
\eea
whereas $y_R \sim {\cal O}(y_c)$. Here $c_{SUSY}$ denotes a combination of flavour off-diagonal elements of the SUSY breaking mass and trilinear parameters between the second and third generation, $N_{ij}$ is the neutralino mixing matrix, and $y_b,y_c$ are the bottom and charm Yukawa couplings. See Refs.\cite{Krizka:2012ah,Grober:2014aha,Grober:2015fia,Hiller:2008wp,Hiller:2009ii,Muhlleitner:2011ww,Belanger:2013oka} for more details.

The partial width of 4BD can be expressed as
$$\Gamma_{4BD}  \sim \frac{g_{\tilde t t \tilde \chi^0}^2}{4!(4 \pi)^{4}} \frac{m_{\tilde t}^7}{m_W^4 m_t^2}$$
when $m_{\tilde t} \gg m_{\tilde \chi^0}$ and as
$$\Gamma_{4BD}  \sim \frac{g_{\tilde t t \tilde \chi^0}^2}{(4 \pi)^{5}} \frac{\Delta m^8}{m_W^4 m_t^2 m_{\tilde t}}$$
for small $\Delta m=m_{\tilde t} -m_{\tilde \chi^0}$~\cite{Delgado:2012eu}. The coupling of the stop to the neutralino and top $g_{\tilde t t \tilde \chi^0}$ depends on the LR admixture of the light stop, and the composition of the neutralino.

These equations show that the 2BD is suppressed if {\it 1.)} the lightest stop is mostly right-handed, {\it 2.)} the flavour off-diagonal elements in the soft-breaking  terms are suppressed, {\it 3.)} the partner of the charm is heavy, and {\it 4.)} the neutralino is a particular linear combination of Bino and Wino. In this case, even the phase space suppressed 4BD could dominate over 2BD. Note also the dependence with $y_b^4$ in the partial width, which introduces a strong dependence on the parameter $\tan \beta$, $\Gamma_{2BD}\propto \tan \beta^4$. 

In summary, the relevant parameters affecting 2BD stop decay are then, besides the level of flavour violation,  the neutralino composition
determined by $\mu$, $M_1$ and $M_2$, the LR stop mixing  and $\tan\beta$. On the other hand, the 4BD does not rely on off-diagonal soft masses, and its dependence on other parameters such as $\mu$ and $\tan \beta$ is different from 2BD.  The full picture of the 2BD versus 4BD interplay is quite complicated in the MSSM multi-dimensional parameter space,
however a judicious choice of parameters~\cite{Krizka:2012ah,Grober:2014aha,Grober:2015fia} allows to see a clearer picture. Indeed, even for $\Delta m$ just above
the  2BD threshold,  the 4BD channel can be dominant if the cancellations discussed above suppress the 2BD. Conversely, for  large  $\tan\beta$ and large values of LR stop mixing,
the 2BD channel can be dominant even for  $\Delta m \simeq M_W$.

One should also mention Ref.~\cite{Belanger:2013oka} where the authors showed that if the neutralino-stop co-annihilation channel is responsible 
for providing the right amount of Dark Matter (DM), then one expects  $\Delta m \simeq 30-40$~GeV, a value (almost) independent of both
stop mass and the stop mixing. This study provides an additional motivation for our choice of $ m_{\tilde{\chi}^\pm_1}>m_{\tilde{t}}$, a choice
which substantially simplifies the study of the LST parameter space. 
At the same time, in the region where $\Delta m < 30-40$~GeV, one expects the
neutralino DM abundance to be consistent with the measured {\it upper limit} of DM relic density.
For  $m_b< \Delta m \lesssim 30-40$~GeV the relative contribution from 2BD and  4BD can be  very different,
such that for any given $\Delta m$,  $\epsilon_{2B}$ takes a value from 0 to 1
depending on $\mu, M_1, M_2$, LR stop mixing  and $\tan\beta$.

In the following, we will perform an analysis making no assumptions on the 
the value of $\epsilon_{2B}$, considering therefore the whole parameter space for the LST scenario.

\section{The setup for the Light Stop analysis}

\subsection{Current Status of the experimental searches.}
The best sensitivity of searches for stops by ATLAS and CMS are reached by focusing on one specific channel, hence assuming a 100\% BR to a final state.  Moreover, the cuts are designed to increase the SUSY signal to SM background ratio in a specific region of SUSY parameter space. As a result, different searches are aimed  to rule out different areas of SUSY parameter space, and these are usually presented in the stop mass ($m_{\tilde{t}}$) vs neutralino mass ($m_{\tilde{\chi}^0_1}$) plane. 
Both ATLAS and CMS have produced summary plots, where they combine all of their stop exclusion results on a single plot, as
 discussed in the Introduction.
Results from both collaborations are very similar, however the ATLAS exclusion limits are slightly more
stringent in the low stop mass region of interest as one can see from  Figure~\ref{stop-atlas-cms}~\cite{ATLAS_Twiki,Aad:2012ywa,Aad:2012xqa,Aad:2012uu,Aad:2014kra,Aad:2014bva,Aad:2014qaa,Aad:2014mfk,Aad:2014nra}. 

Even under these stringent assumptions, we clearly see that there are areas of parameter space which still allow light stops. For example, if the neutralino mass were $m_{\tilde{\chi}^0_1} \gsim 240$ GeV, then any stop mass down to around 280 GeV would be still allowed. If $m_{\tilde{\chi}^0_1} \lsim 240$ GeV,  then stops as light as 110 GeV may be allowed depending on the mass gap, $\Delta m$, between the stop and neutralino. The two main regions which are not excluded even for these very light stops are where $\Delta m$ is around $M_W$, and where $\Delta m \approx m_t$. In both these regions, the stop decays to an on-shell $W$-boson or top quark, with very little energy for the neutralino. Therefore there is very little Missing Transverse Energy (MET) from the undetected neutralino, which makes it difficult to distinguish signal from the large SM background.  

Our goal here is two-fold. First we wish to extend the ATLAS analysis into the region with light stops where $\Delta m$ is slightly larger than $M_W$, with the intention of ruling out the lowest mass stop regions which are presently still allowed experimentally. We specifically choose to extend the ATLAS bounds because in this region they are more stringent than the corresponding CMS results. Secondly, by reproducing the analyses ourselves, and validating them against the published experimental results, we will have the freedom to alter branching ratios, allowing us to explore the consequences on the exclusion limits of a more realistic model where the stop has more than one decay channel with a significant branching ratio.

To overcome the problem related to the small mass gaps between the stop and neutralino, resulting in  little momentum
release,
one can use events with a high-$p_T$ initial (and/or final)  state gluon or quark radiation - ISR (FSR) -
which would recoil against  the $\tilde{t}\bar{\tilde{t}}$ pair, leading to a larger boost of the neutralinos from stops decays. The resulting signature is a high-$p_T$ jet  and the high missing energy (MET)
from recoiled neutralinos against this jet.  All of the ATLAS searches in the $\Delta m < M_W$ region are monojet searches, with cuts for a high-$p_T$ jet and high MET.

As stated previously, in this paper we extend the ATLAS exclusion into the $\Delta m > M_W$ region. In particular, the intention is to extend the regions in Refs.~\cite{Aad:2014nra} and \cite{Aad:2014kra} (the salmon coloured and dark grey regions in Figure~\ref{stop-atlas-cms}), which both rule out a large region where $\Delta m < M_W$, but are artificially cut off at around the $\Delta m = M_W$ line, where it looks likely they could have been extend further. 

In Ref.~\cite{Aad:2014nra} where a $\tilde{t}_1 \to \tilde{\chi}^0_1 c$ decay is assumed, it is stated that the maximum $\Delta m$ considered is 82 GeV. No further explanation is given, however its likely this is in part due to the fact that if one would assume no tree level flavour violation, then the region of parameter space where the branching ratio to $\tilde{t}_1 \to \tilde{\chi}^0_1 c$ is 100\% becomes very small as $\Delta m$ becomes much larger than this. However, as discussed in Section~2, sizeable $\tilde{t}_1 \to \tilde{\chi}^0_1 c$ BRs are still possible for mass gaps up to at least $\Delta m \approx 110$ GeV when flavour violation within experimental limits are allowed, and it is important to exclude this experimentally.

In Ref.~\cite{Aad:2014kra} on the other hand, where a $\tilde{t}_1 \to b f f^{\prime} \tilde{\chi}^0_1$ decay is assumed, they state that ``{\it generating the full event with MadGraph would be computationally too expensive.''}. As a result, their $\tilde{t}$ are decayed using Pythia, which produces isotropic decays. This will not be valid when the W bosons are on shell. This seems to be at least part of the reason the results have been restricted to $\Delta m < 80$ GeV (which is not explicitly stated). As these omissions are both important and possible to rectify, these are the analyses we extend in this chapter.

\subsection{Tools and Framework for Analysis}
In order to extend these results, we reproduced the signal samples and analysis for three ATLAS analysis~\cite{Aad:2014nra,Aad:2014kra} which we will call; (i) monojet analysis, (ii) monojet with $c$-tagging analysis, and (iii) monojet with 1 lepton analysis. They are discussed in the following subsections.

\subsubsection{Monojet, $\tilde{t} \to \tilde{\chi}^0_1 c$}
This analysis is described in \cite{Aad:2014nra}. It assumes a 100\% branching ratio to $\tilde{t} \to \tilde{\chi}^0_1 c$ and its main aim is to rule out the very small $\Delta m$ region where the $c$-jets from the decay will usually be too soft to identify (roughly $\Delta m < 30$ GeV although ATLAS do not give a value). Therefore in monojet events, the signature will be one high-$p_T$ jet and a large $E_T^{\rm miss}$, with a small number of soft jets. 

First, the events undergo a pre-selection, requiring an $E_T^{miss} > 150$ GeV, at least one jet with a $p_T > 150$ GeV and $|\eta|<2.8$, and vetoing any event with a muon with $p_T > 10$ GeV or an electron with $p_T > 20$ GeV. Following this, as a result of the softness of the decay products, a maximum of three jets with $p_T > 30$ GeV and $|\eta| < 2.8$ are allowed. Additionally, the azimuthal separation between the missing transverse momentum direction and that of each jet has a minimum bound, $\Delta \phi(jet,p_T^{\rm miss}) > 0.4$, which ATLAS used to reduce the multijet background where the large $E_T^{\rm miss}$ originates mainly from jet energy mismeasurement. In order to optimise the search reach, 3 separate signal regions were defined (denoted M1, M2, M3), with increasing minimum thresholds for $p_T$ and $E_T^{\rm miss}$ to exclude increasing stop and neutralino masses. For the M1, M2 and M3 regions, the thresholds are respectively $p_T > 280$ GeV, $E_T^{\rm miss}> 220$ GeV for M1, $p_T > 340$ GeV, $E_T^{\rm miss}> 340$ GeV for M2, and $p_T > 450$ GeV, $E_T^{\rm miss}> 450$ GeV for M3. These selection cuts are summarised in Table~\ref{Tab:Monojet}.

\begin{table}[!h]
\begin{center}
\begin{tabular}{|l|c|c|c|}
\hline
\multicolumn{4}{|c|}{\textbf{Monojet Search}} \\ \hline
\multicolumn{4}{|c|}{{\it Applied to all 3 signal regions (M1, M2, M3)}}\\
\multicolumn{4}{|l|}{At most 3 jets with $p_T > 30$ GeV and $|\eta| < 2.8$}\\
\multicolumn{4}{|l|}{$\Delta \phi(jet,p_T^{\rm miss}) > 0.4$}\\
\hline
\textbf{Signal region} \hspace{20mm} &  \hspace{5mm} \textbf{M1} \hspace{5mm} & \hspace{5mm} \textbf{M2} \hspace{5mm} & \hspace{5mm} \textbf{M3} \hspace{5mm} \\
Minimum leading jet $p_T$ (GeV)  & 280 & 340 & 450\\
Minimum $E_T^{\rm miss}$ (GeV) & 220 & 340 & 450\\
\hline 
\end{tabular}
\caption{Analysis cuts for the pure monojet search in the $\tilde{t} \to \tilde{\chi}^0_1 c$ channel. There are 3 separate signal regions, M1, M2 and M3. The cuts applied to all 3 regions are in the top row, with the signal region dependent cuts in the lower row.  \label{Tab:Monojet}}
\end{center}
\end{table}

The SUSY signal samples were produced at leading order using {\tt MadGraph5}~\cite{Maltoni:2002qb,madgraph:2004sl,Alwall:2011uj} with a CTEQ6L1 PDF, with the cross section rescaled using a K-factor calculated with next-to-leading order (NLO) supersymmetric QCD corrections and the resummation of soft gluon emission at next-to-leading-logarithmic (NLL) accuracy using the NLL-fast computer program~\cite{Beenakker:2011fu,Beenakker:2010nq,Beenakker:1997ut}. In view of the fact that for large $\Delta m$, the veto of any event with a fourth jet with $p_T > 30$ GeV can reduce the selection efficiency by around 50\%, and that this can be from a second initial state radiation jet (with the $2^{\rm nd}$ and $3^{\rm rd}$ highest-$p_T$ jets from the $c$-quarks from stop decays), two-jet matching using the $k_T$-jet MLM scheme~\cite{Alwall:2007fs} was used to ensure accuracy of the $p_T$ of subleading ISR jets. The showering is done using {\tt Pythia-6}~\cite{PYTHIA,PYTHIA_interface,Sjostrand:2006za} and the detector simulation using {\tt Delphes-3}~\cite{Cacciari:2005hq,Cacciari:2011ma,deFavereau:2013fsa}.
The subsequent analysis and application of cuts was conducted using the {\tt ROOT} Data Analysis Framework~\cite{Brun:1997pa}. Each point in the $m_{\tilde{t}}$ vs $m_{\tilde{\chi}^0_1}$ plane was ruled out if for any of the signal regions (M1, M2, M3), the cross section of the signal sample and the efficiencies of the selection cuts predicted a larger number of signal events than the 95\% confidence limits (CL) upper limit on BSM events which is provided by ATLAS in the paper.

\subsubsection{Monojet with c-tagging, $\tilde{t} \to \tilde{\chi}^0_1 c$}
This analysis is also described in Ref.~\cite{Aad:2014nra}. It again assumes a 100\% branching ratio to $\tilde{t} \to \tilde{\chi}^0_1 c$, and its main purpose is to rule out the region with a larger but still relatively small $\Delta m$, (roughly 30 GeV $< \Delta m < 80$ GeV although ATLAS does not give a value), where the c-jets from the decay will usually be hard enough to identify, but softer than the initial state radiation. Therefore the signature will be relatively large multiplicity jets with a charm jet as one of the subleading jets.

At ATLAS, the $c$-tagging is implemented via a dedicated algorithm using multivariate techniques which combine information from the impact parameters of displaced tracks and topological properties of secondary and tertiary decay vertices reconstructed within the jet. For this study, they used two operating points for the $c$-tagging called the {\it medium} and {\it loose} operating points. The {\it medium} operating point has a $c$-tagging efficiency of $\approx 20$\%, and a rejection factor of $\approx 8$ for $b$-jets, $\approx 200$ for light-flavour jets, and $\approx 10$ for $\tau$-jets, while the {\it loose} operating point has a $c$-tagging efficiency of $\approx 95$\%, with a rejection factor of $\approx 2.5$ for $b$-jets, but no significant rejection of light-flavour or $\tau$-jets. For our analysis, we used these quoted efficiencies and rejection factors, as well as representative data-to-simulation multiplicative scale factors given in the ATLAS paper~\cite{Aad:2014nra} of 0.9 for simulated heavy-flavour tagging and 1.5 for mistagging of light-jets as charm jets.

Once more, the events undergo a pre-selection (slightly different to the monojet pre-selection), requiring an $E_T^{miss} > 150$ GeV, at least one jet with $p_T > 150$ GeV and $|\eta|<2.5$, and vetoing any event with a muon or electron with $p_T > 10$ GeV.

\begin{table}[!h]
\begin{center}
\begin{tabular}{|l|c|c|}
\hline
\multicolumn{3}{|c|}{\textbf{Monojet with $c$-tagging Search}} \\ \hline
\multicolumn{3}{|c|}{{\it Applied to both signal regions (C1, C2)}}\\
\multicolumn{3}{|l|}{At least four jets with $p_T > 30$ GeV and $|\eta| < 2.5$}\\
\multicolumn{3}{|l|}{$\Delta \phi(jet,p_T^{\rm miss}) > 0.4$}\\ 
\multicolumn{3}{|l|}{All four jets must pass loose tag requirements ($b$-jet vetoes)}\\
\multicolumn{3}{|l|}{At least one medium charm tag in the three subleading jets}\\ 
\hline
\textbf{Signal region} \hspace{35mm} &  \hspace{7mm} \textbf{C1} \hspace{7mm} & \hspace{7mm} \textbf{C2} \hspace{7mm}   \\
Minimum leading jet $p_T$ (GeV)  & 290 & 290  \\
Minimum $E_T^{\rm miss}$ (GeV) & 250 & 350  \\
\hline 
\end{tabular}
\caption{Analysis cuts for the monojet with $c$-tagging search in the $\tilde{t} \to \tilde{\chi}^0_1 c$ channel. There are 2 separate signal regions, C1 and C2. The cuts applied to both regions are in the top row, with the signal region dependent cuts in the lower row. \label{Tab:Ctag}}
\end{center}
\end{table}
Following this, due to the likelihood of multiple jets, a minimum of four jets with $p_T > 30$ GeV and $|\eta| < 2.5$ and $\Delta \phi(jet,p_T^{\rm miss}) > 0.4$ are required. Additionally, there is a veto against any event containing $b$-jets by using a loose $c$-tag requirement, and a requirement that at least one of the three subleading jets passes a medium $c$-tag. Again in order to optimise the search reach, 2 separate signal regions were defined (denoted C1 and  C2), both requiring their leading jet to have $p_T > 290$ GeV, but with C1 requiring $E_T^{\rm miss} > 250$ GeV and C2 requiring $E_T^{\rm miss} > 350$. These selection cuts are summarised in Table~\ref{Tab:Ctag}. 

Once more, the SUSY signal samples were produced using {\tt Madgraph5} (with 2-jet matching and a CTEQ6L1 PDF), {\tt Pythia} and {\tt Delphes-3}, with subsequent analysis conducted using {\tt ROOT}. Each point in the $m_{\tilde{t}}$ vs $m_{\tilde{\chi}^0_1}$ plane was ruled out if for any of the signal regions (C1, C2) predicted a larger number of signal events than the 95\% CL upper limit on BSM events provided by ATLAS.

\subsubsection{Monojet with 1 lepton, $\tilde{t} \to b f f^{\prime} \tilde{\chi}^0_1$}
This analysis is described in Ref.~\cite{Aad:2014kra}. It assumes a 100\% branching ratio to $\tilde{t} \to b f f^{\prime} \tilde{\chi}^0_1$. Like the previous 2 analysis discussed above, it is separated into 2 signal regions, with the first, labelled bCa\_low, aiming to probe mass scenarios where $\Delta m < 50$ GeV, and the second, bCa\_med, intended to probe 50 GeV $< \Delta m < 80$ GeV. 

\begin{table}[!h]
\begin{center}
\begin{tabular}{|l|c|c|}
\hline
                                   &     \hspace{12mm}   \textbf{bCa\_low}  \hspace{12mm}   &     \hspace{12mm}    \textbf{bCa\_med} \hspace{12mm}              \\  \hline
\multirow{2}{*}{\textbf{Lepton} \hspace{12mm}}   & \multicolumn{2}{|c|}{7 GeV $< p_T^{\rm electron} < 25$ GeV } \\ 
                                   & \multicolumn{2}{|c|}{6 GeV $< p_T^{\rm muon} < 25$ GeV }     \\ \hline
\multirow{2}{*}{\textbf{Jets}}     &   $\ge 2$ with       &         $\ge 3$ with                 \\ 
                                   &  $p_T > 180,25$ GeV &  $p_T > 180,25,25$ GeV              \\ \hline
\textbf{b-tagging}                 &  \multicolumn{2}{|c|}{ $\ge 1$ sub-leading jet b-tagged (70\% eff.) }\\ \hline
\textbf{b-veto}                    &  \multicolumn{2}{|c|}{ $1^{\rm st}$ jet not b-tagged (70\% eff.) }\\ \hline
$\mathbf{E_T^{\rm miss}}$            &      $> 370$ GeV    &    $> 300$ GeV                      \\ \hline
$\mathbf{E_T^{\rm miss}/m_{\rm eff}}$& $> 0.35$    &    $> 0.3$                           \\ \hline
$\mathbf{m_T}$                     &      $> 90$ GeV     &    $> 100$ GeV                       \\ \hline
\end{tabular}
\caption{Analysis cuts for the monojet with 1-lepton search in the $\tilde{t} \to b f f^{\prime} \tilde{\chi}^0_1$ channel. There are 2 separate signal regions, bCa\_low and bCa\_med.\label{Tab:1lepton}}
\end{center}
\end{table}
There are a number of differences between the event selection criteria for the 2 signal regions, all of which are presented in Table~\ref{Tab:1lepton} for convenience. For bCa\_med there is a requirement for $\ge 3$ jets to suppress the SM $W$+jets background, while for bCa\_low this is lowered to $\ge 2$ to avoid large acceptance losses. $m_{eff}$ is defined by 

\begin{equation}
m_{eff} = H_T + p_T^l + E_T^{\rm miss}
\end{equation}

\noindent where $H_T$ is the scalar $p_T$ sum of the four leading jets and $p_T^l$ is the $p_T$ of the single charged lepton in the event. Assuming the lepton mass is negligible, the transverse mass ($m_T$) is defined by,

\begin{equation}
m_T = \sqrt{2.p_T^l.E_T^{\rm miss} \left( 1-\cos \Delta \phi(\vec{l},\vec{p}_T^{\rm miss}) \right)}.
\end{equation}

\noindent Here $\Delta \phi(\vec{l},\vec{p}_T^{\rm miss})$ is the azimuthal angle between the lepton momentum and the $\vec{p}_T^{\rm miss}$ directions.

This is the analysis which ATLAS deemed computationally too expensive to produce the full matrix element for the SUSY signal sample, instead using Pythia which decays the $\tilde{t}_1$ isotropically. This limits the analysis to $\Delta m < 80$ GeV and fails to rule out the region we're interested in. Without flavour violation (beyond the SM), the assumption of a 100\% branching ratio to $\tilde{t} \to b f f^{\prime} \tilde{\chi}^0_1$ is correct for most of parameter space when $\Delta m > 80$, and it would be particularly useful to extend this analysis into this space. 

We used {\tt Madgraph5} to produce the signal events. This was impossible to do accurately until November 2014, due to a bug in MadGraph which was fixed between Version-2.2.1 and Version-2.2.2\footnote{For small $\Delta m \lsim 80$ GeV, the bug resulted in MadGraph incorrectly including many of the soft jets from stop decays in the matching scheme, with the result of a large proportion of the events being incorrectly vetoed, giving cross sections far smaller than their correct values.}. Once this bug was fixed, the generation of events was computationally intensive but achievable. Jet matching was required due to the added complication in this region that after selection cuts, the leading jet is sometimes from the decay products rather than being initial state radiation (ISR), which in the absence of matching leads to an infrared divergence of the ISR. Again the PDF used was CTEQ6L1 PDF, with {\tt Pythia}, {\tt  Delphes-3} and {\tt ROOT} used for the rest of the signal generation and analysis. Each point in the $m_{\tilde{t}}$ vs $m_{\tilde{\chi}^0_1}$ plane was ruled out if for any of the signal regions (bCa\_low, bCa\_med) we predicted a larger number of signal events than the 95\% CL upper limit on BSM events provided by ATLAS.

\section{Results on the exclusion of the LST parameter space}

\subsection{2-body   $\tilde{t} \to \tilde{\chi}^0_1 c$ decay channel analysis:
monojet signature without and with c-tagging}

\begin{figure}[htb]
\centering
\includegraphics[width=0.8\textwidth]{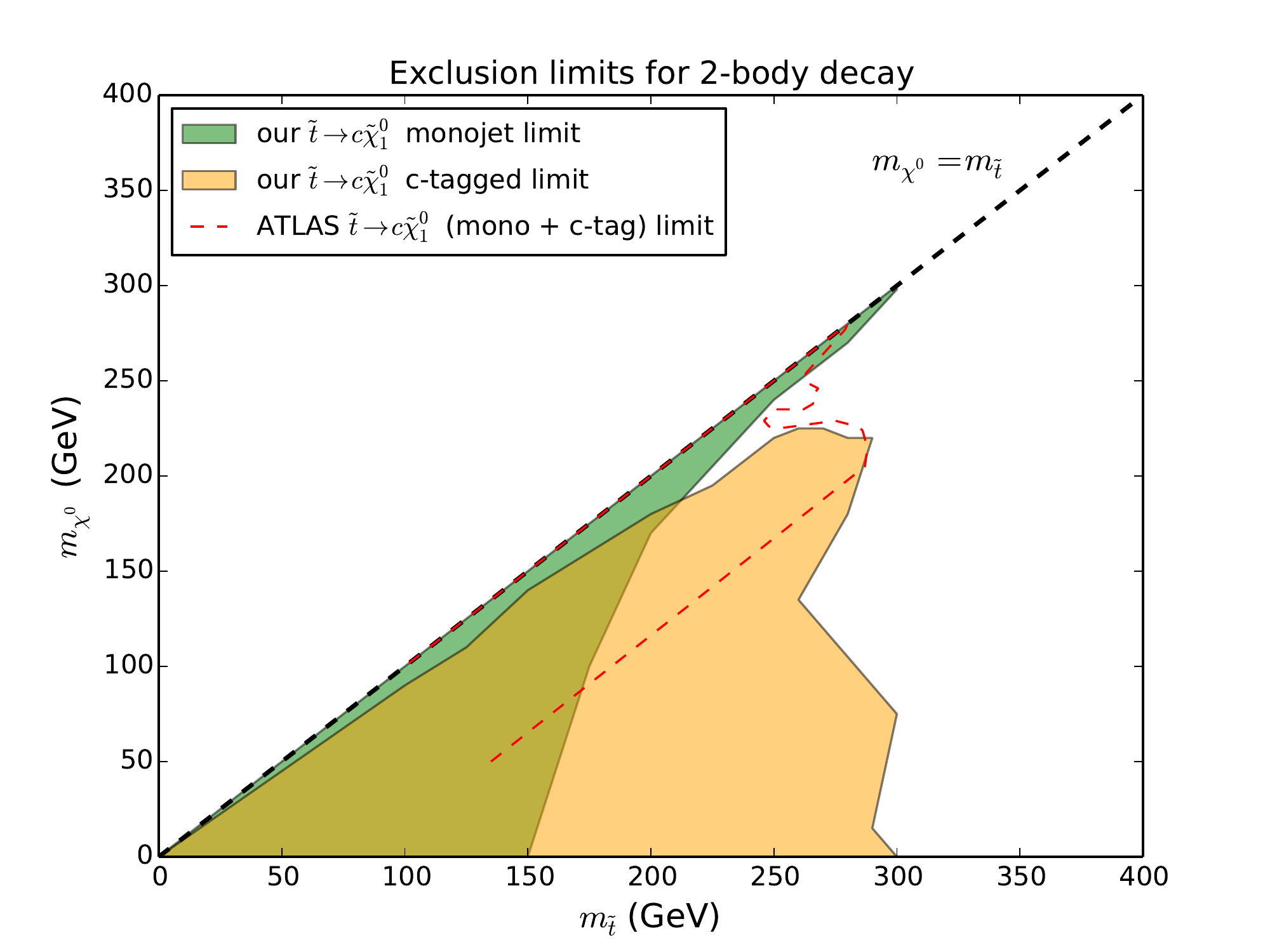}
\caption{The green region denotes the region excluded by the monojet analysis while the orange region is excluded by the monojet with $c$-tagging analysis. For both the exclusion is at a 95\% CL assuming a 100\% branching ratio to $\tilde{t} \to \tilde{\chi}^0_1 c$. The dashed red line is an outline of the region excluded by ATLAS after they conducted and combined the same two analyses.
\label{fig:res_mono}}
\end{figure}

By applying the cuts required to define these signal regions as described in the previous 
section, and utilising confidence level limits on the number of excess events 
over the background obtained by ATLAS and supplied in their paper \cite{Aad:2014nra}, we have 
found the exclusion region from these channels in the ($m_{\tilde{\chi}^0_1}-m_{\tilde{top}_1}$)
parameter space.  The 95\%CL exclusion region for 
the $\tilde{\chi}^0_1 c$ decay channel from analyses looking for a monojet signature (without c-tagging)
is presented in Fig.~\ref{fig:res_mono} and indicated by the green colour.

The ATLAS collaboration only presents their result after combining this exclusion region with that of the monojet with $c$-tagging search, with their combined exclusion being a salmon pink colour in Fig.~\ref{stop-atlas-cms}(top). This is done because this combination gives the entire region ruled out given the assumption that $\tilde{t} \to \tilde{\chi}^0_1 c$ is the only decay channel. It is this combined region's outline that is given by the red dashed line in Figure~\ref{fig:res_mono}. Our monojet result reproduces the wedge shape seen in the ATLAS exclusion near $m_{\tilde{t}} = m_{\tilde{\chi}^0} \approx 270$ GeV. 

One should stress that we have extended our analysis into the $\Delta m \gtrsim 80$ GeV region:
one can see that for $m_{\tilde{t}} \lsim 170$ GeV, 
a new region beyond $\Delta m < 80$ GeV  is ruled out which is not covered by the ATLAS analyses.
If we assume that the decay is entirely via $\tilde{t} \to \tilde{\chi}^0_1 c$, this monojet analysis alone rules out stops with $m_t < 150$ GeV.

%\subsubsection{Monojet with c-tagging Analysis}

In Fig.~\ref{fig:res_mono} we also present
results for the 95\% exclusion region for the monojet with $c$-tagging  denoted by the orange region. 
Firstly, we see that we have successfully recreated the ``bulge'' in the ATLAS results, where 40 GeV $< \Delta m <$ 80 GeV and $m_{\tilde{t}} \approx 270$ GeV. When this is combined with the green monojet exclusion, we find that other than a small wedge when $m_{\tilde{t}} \approx 240$ GeV, $m_{\tilde{\chi}^0} \approx 210$ GeV, we agree well with ATLAS for the masses for which they have produced results, as we should expect. This agreement validates our signal sample generation and analysis.
Secondly, our 95\% CL extends well beyond the region excluded by ATLAS, all the way down to massless neutralinos. This means that if the assumption that $\tilde{t} \to \tilde{\chi}^0_1 c$ has a BR of 100\% were true, light stops are ruled out for $m_{\tilde{t}} < 210$ GeV regardless of neutralino mass. As discussed in section~\ref{sec:StopDecays}, the BR for this decay can vary a lot in the LST parameter space, so the 
assumption of a 100\% BR over the entire region should be considered as a convenient way to present the results
and not a realistic physics picture, as we discuss in detail below.

\subsection{4-body,  $\tilde{t} \to b f f^{\prime} \tilde{\chi}^0_1$ analysis: monojet with 1 lepton signature}
\begin{figure}[htb]
\centering
\includegraphics[width=0.9\textwidth]{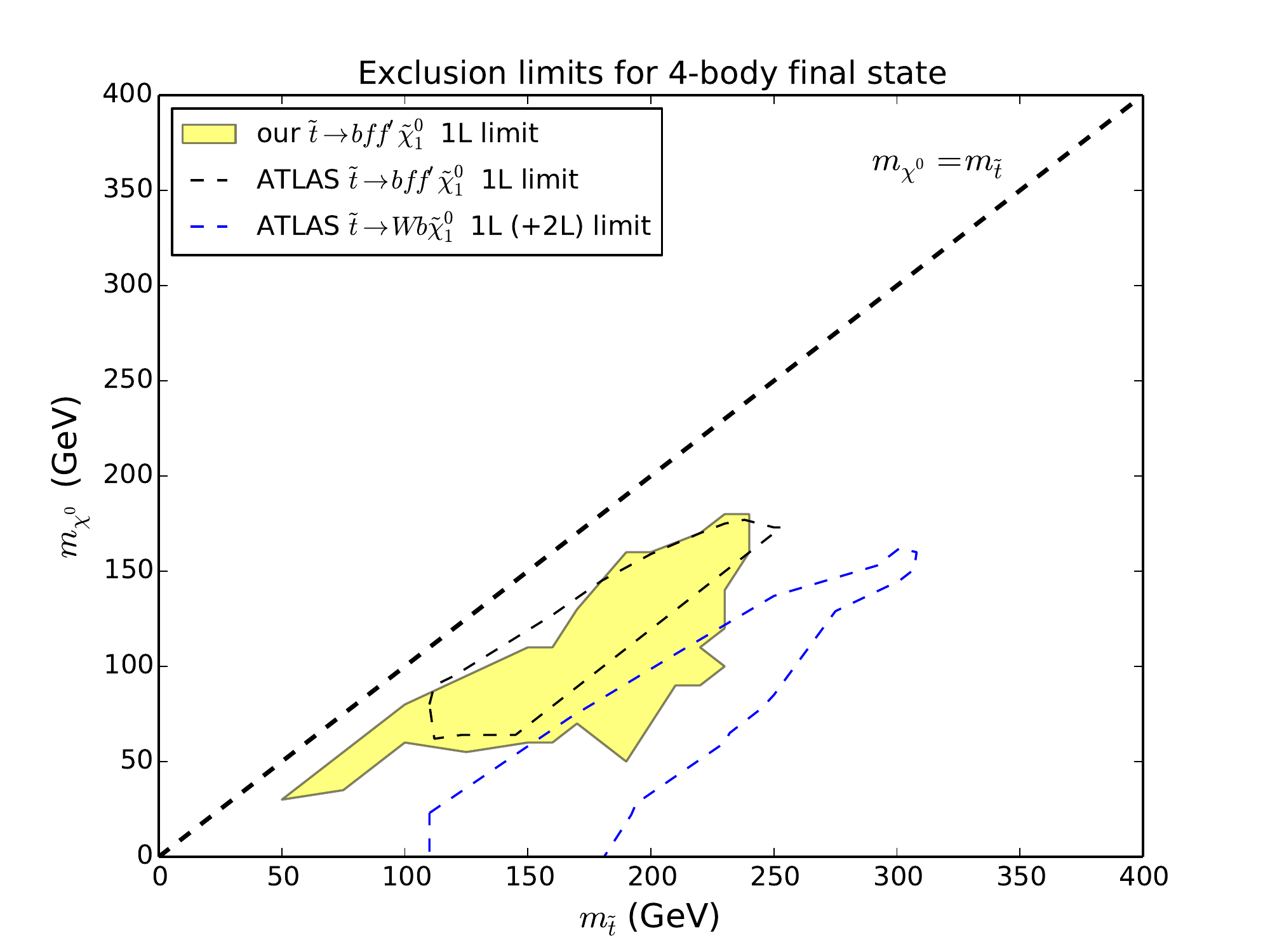}\caption{The yellow area is excluded at a 95\% CL by the monojet with 1-lepton analysis, assuming a 100\% branching ratio to $\tilde{t} \to b f f^{\prime} \tilde{\chi}^0_1$. The dashed black line is an outline of the region excluded by ATLAS for the same analysis. The region inside the dashed blue line is excluded by a a combination of 2 different ATLAS analyses which also assume a $\tilde{t} \to b f f^{\prime} \tilde{\chi}^0_1$ decay.
\label{fig:1lepton}}
\end{figure}

The results of this analysis is presented separately from the monojet and monojet with charm tagging results as the assumed decay process and the respective signature are different. In Figure~\ref{fig:1lepton}, we show our 95\% CL excluded region, compared to the analogous ATLAS result outlined in dashed black. Also included on the plot is another ATLAS analysis in dashed blue
(1- and 2-lepton analysis which are different from ours) which we did not attempt to reproduce as we have no reason to believe that we could extend it.
This is included to make it visually clear which region we particularly intended to rule out; the region {\it between} the two ATLAS exclusions.

Our exclusion region once more agrees reasonably well with ATLAS for $\Delta m < 80$ GeV, acting as a validation for our methods. However it also extends beyond this bound filling the previously unexcluded gap between the two ATLAS analyses, where $\Delta m$ is slightly larger than $M_W$. Therefore, under the assumption that stops only have a 4-body decay, we have successfully ruled out a large part of the remaining phase space for light stops with masses of around 150 GeV $< m_{\tilde{t}} <$ 200 GeV. As discussed in Section~\ref{sec:StopDecays}, if there is no flavor violation (beyond the SM) in the MSSM, then stops exclusively decaying to 4-body is a reasonable assumption over much of parameter space. More generally however 2-body decays can also occur in this mass range.

%%%%%%%%%%%%%%%%%%%%%%%%%%%
%%%%%%%%%%%%%%%%%%%%%%%%%%%%
\subsection{Combining new  and existing results}
In this section, to see the full region in the $m_{\tilde{t}}$ vs $m_{\tilde{\chi}^0_1}$ plane which is now excluded, we combine our results with those of ATLAS, including ATLAS analyses which we did not attempt to reproduce. As previously, it is sensible to consider the two decay channels separately which we do below.

\subsubsection{2-body, $\tilde{t} \to \tilde{\chi}^0_1 c$}
\begin{figure}[htb]
\centering
\includegraphics[width=0.9\textwidth]{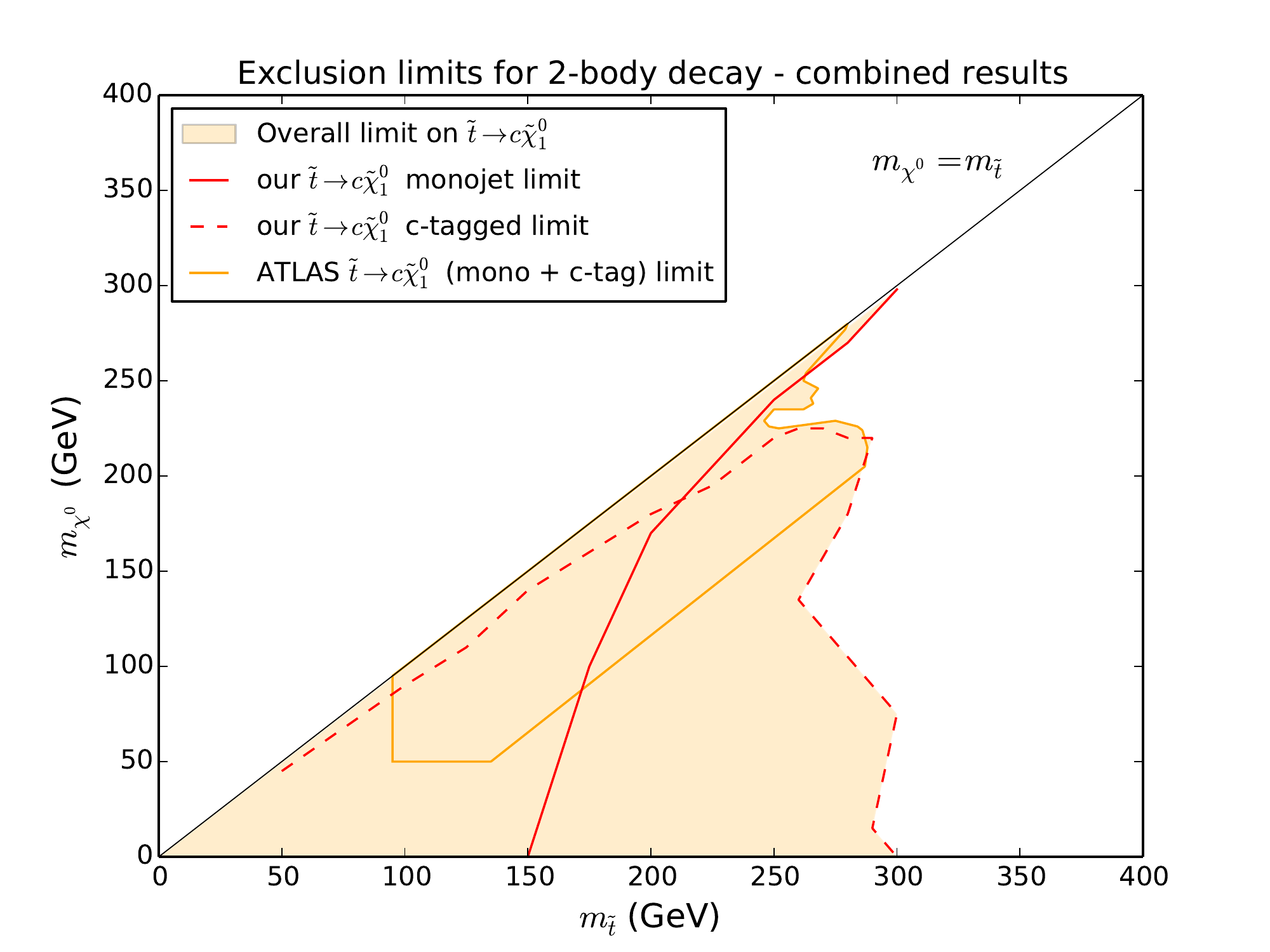}
\caption{The area shaded yellow in the $m_{\tilde{t}}$ vs $m_{\tilde{\chi}^0}$ plane has been excluded at the 95\% CL after both our results and ATLAS results are included. The red outlines show regions excluded by our analysis (solid: monojet analysis, dashed: monojet with $c$-tagging analysis). The solid orange line outlines the region excluded by ATLAS.
\label{fig:2body_exc}}
\end{figure}

As we have reproduced all of the ATLAS analyses which assume a $\tilde{t} \to \tilde{\chi}^0_1 c$ decay, combining our results with that of ATLAS only excludes an additional small wedge shaped region around $m_{\tilde{t}} \approx 240$, $m_{\tilde{\chi}^0} \approx 210$. These combined results are shown in Figure~\ref{fig:2body_exc}.

Our conclusion here is very similar to that prior to combining our results with ATLAS, but with the lower bound on the stop mass increased to around 240 GeV. If true, as having $m_{\tilde{t}} < m_t$ is a necessary condition for the light stop scenario of EW baryogenesis, this scenario would have been ruled out, but as this decay is disfavoured for moderate and large values of $\Delta m$ this conclusion is invalid more generally.

\subsubsection{Four Body, $\tilde{t} \to b f f^{\prime} \tilde{\chi}^0_1$}
\begin{figure}[!htb]
\centering
\includegraphics[width=0.9\textwidth]{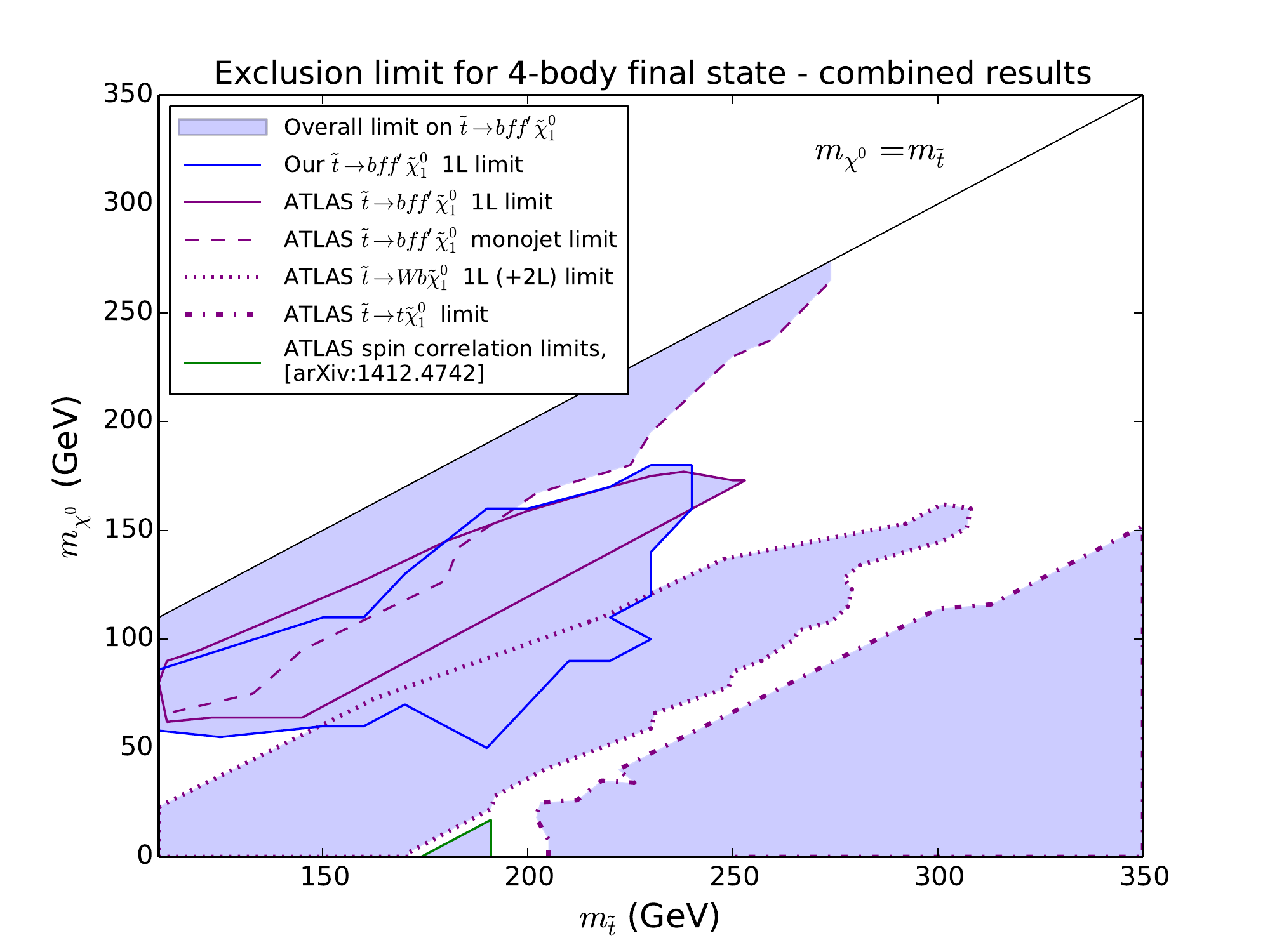}
\caption{Area in the $m_{\tilde{t}}$ vs $m_{\tilde{\chi}^0}$ plane which has been excluded at the 95\% CL after combining our results with ATLAS. Blue outline - excluded by our monojet with lepton search. Purple outlines - regions excluded by ATLAS searches. Green outline - Excluded by ATLAS search via top-antitop spin correlations.
\label{fig:4body_exc}}
\end{figure}

In Figure~\ref{fig:4body_exc} we combine our results for the monojet with one lepton analysis with all of the ATLAS analyses which assume the same $b f f^{\prime} \tilde{\chi}^0_1$ final state. 
The total area excluded at the 95\% CL is shaded in blue. The outline of our contribution to the total exclusion area has a solid blue line, while all of the ATLAS exclusion results are outlined in purple. The only exception is another ATLAS study which is based on top-antitop spin correlations, whose outline is green.

The goal of the study was to rule out as much of the region with light stops as possible, in particular where $\Delta m$ is slightly larger than $M_W$ as this is where extending the ATLAS 95\% CL was most likely to be successful. Inspection of Figure~\ref{fig:4body_exc} shows that the addition of our analysis to ATLAS's results has  achieved this, closing much of this remaining region and bridging the gap between ATLAS's $\tilde{t} \to b f f^{\prime} \tilde{\chi}^0_1$ analyses (where $\Delta m < 80$ GeV) and $\tilde{t} \to b W \tilde{\chi}^0_1$ analyses (where $\Delta m \gtrsim 80$ GeV).

However, there still remains a small area where 100 GeV $\lsim m_{\tilde{t}} \lsim 140$ GeV and 25 GeV $\lsim m_{\tilde{\chi}^0} \lsim$ 50 GeV where light stops are still allowed, as well as a narrow band along the $\Delta m \approx m_t$ line, and a small region where 191 GeV $< m_{\tilde{t}} \lsim 205$ GeV near where the neutralino is massless. Therefore even with the assumption of a 4-body decay BR of 100\% there remains a small region where the stop is still light enough to allow EWBG. Furthermore as discussed in Section~\ref{sec:StopDecays}, this assumption of exclusively 4-body decays is not valid as 2-body decays are able to occur for $\Delta m$ at least up to 110 GeV when FV is allowed. 

There are ongoing efforts to reduce the region where $\Delta m \approx m_t$ further. These include spin correlation approaches\cite{Han:2013lna}, and methods where the stop manifests as a disagreement between the theoretical and experimental values of the top cross section\cite{Czakon:2014fka}.

%%%%%%%%%%%%%%%%%%%%%%%%%%%%%%%%%%%%
%%%%%%%%%%%%%%%%%%%%%%%%%%%%%%%%%%%%
\subsection{Model independent results for various  branching ratios in generic LST parameter space}
Thus far all the results presented assume a 100\% branching fraction, either decaying via $\tilde{t} \to \tilde{\chi}^0_1 c$ or $\tilde{t} \to b f f^{\prime} \tilde{\chi}^0_1$. These results are convenient for presentation purposes, however 
if the  the LST scenario is realised in nature, we need to consider a more realistic scenario, and study the
parmater space for various values of BR.
This is what we do in this section, where we allow intermediate values of BR, assuming that these are the only two decay channels such that their branching ratios add to 100\%, which is the definition of the LST parameter space which we study here.

The procedure followed was a simple procedure of adjusting the cross section and therefore the number of predicted signal events, according to the branching ratios. A point in the mass plane is excluded if the number of signal events in either channel was predicted to be larger than the 95\% confidence limits (CL) upper limit on BSM events provided by ATLAS. This naive method is likely to be more pessimistic than a more sophisticated likelihood contour method. Furthermore, to produce these results, we have also assumed that any event with a mixed decay, i.e. where the two stops which are pair-produced decay to one of each of the two different final states, will not pass the selection cuts. While the efficiency of such events is likely to be low, this assumption is unlikely to be true for every mixed event, and therefore the exclusion regions presented here should be considered a {\it minimum} exclusion region.

\begin{figure}[htb]
\centering
\vspace*{-15mm}
\makebox[\textwidth][c]{
  \includegraphics[width=0.5\textwidth]{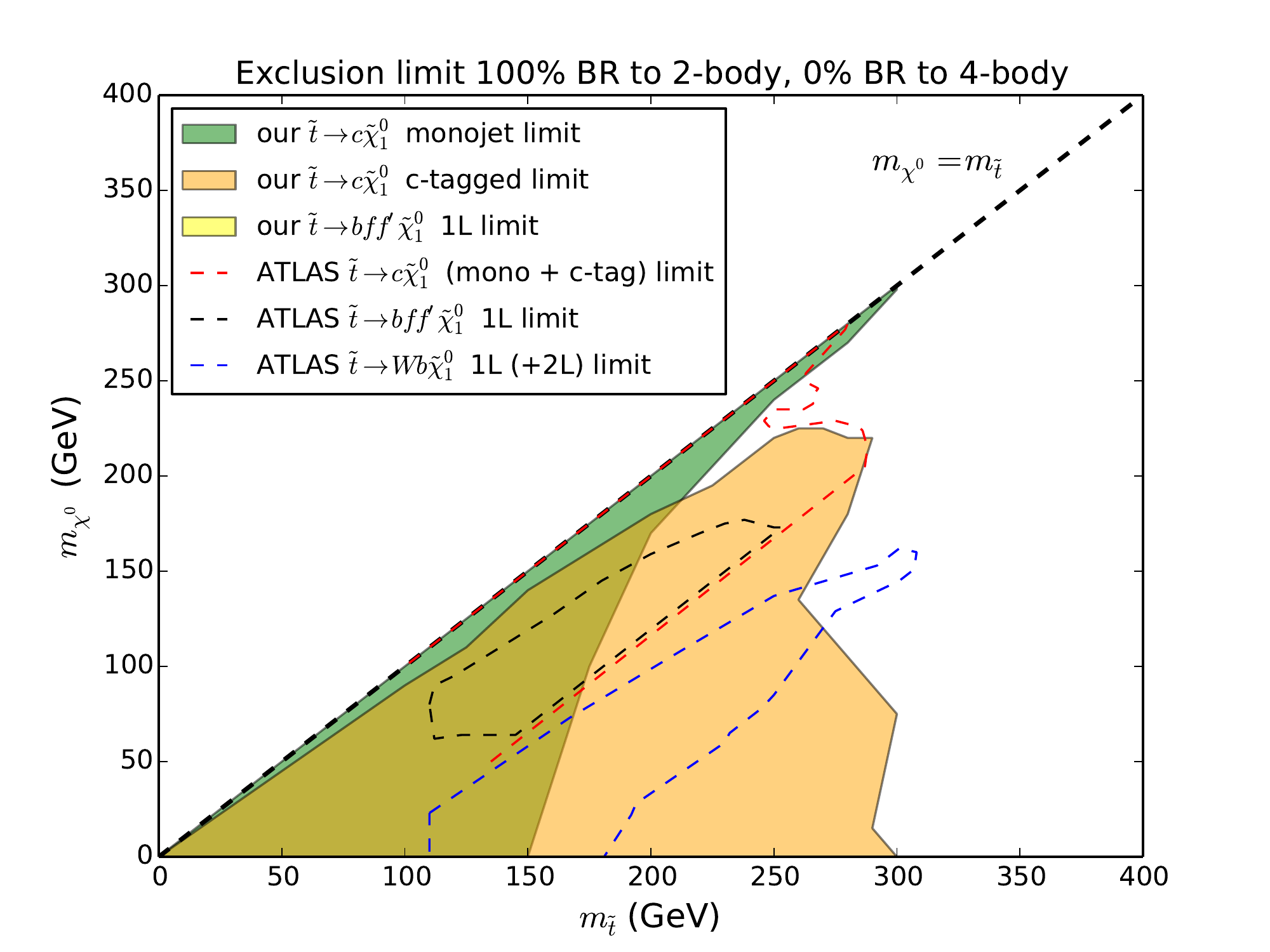}
  \includegraphics[width=0.5\textwidth]{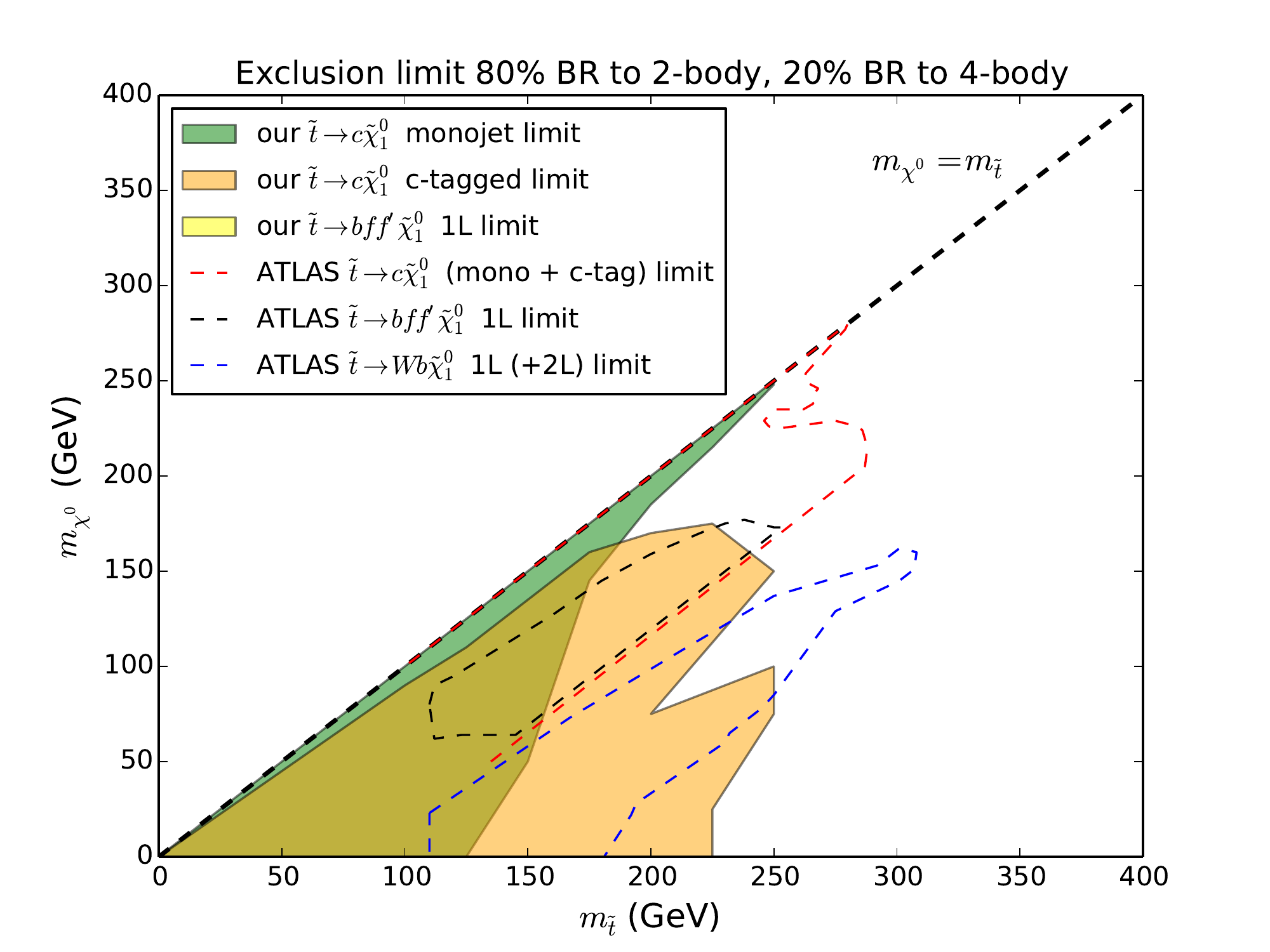}}\\
\makebox[\textwidth][c]{
  \includegraphics[width=0.5\textwidth]{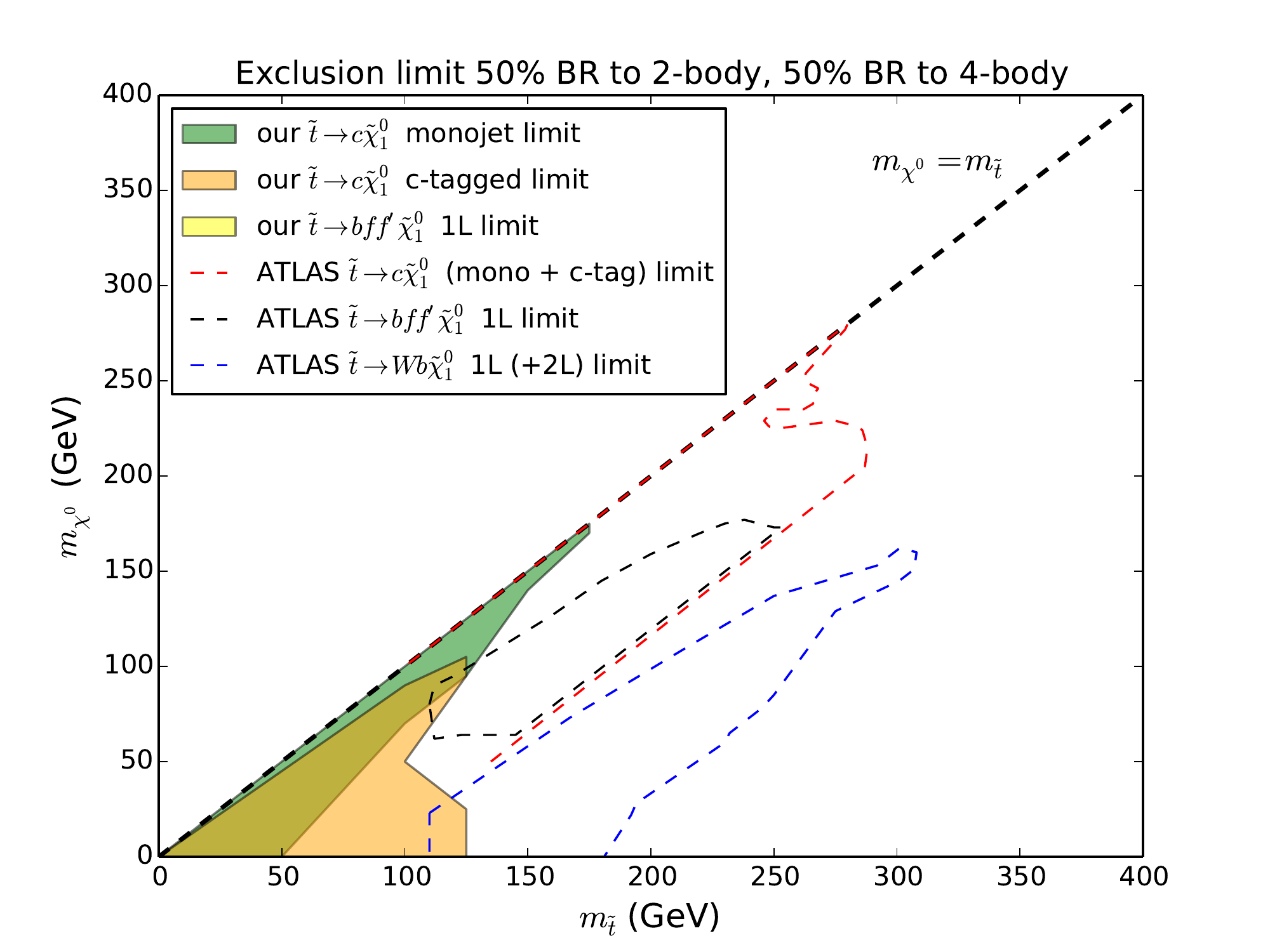}
  \includegraphics[width=0.5\textwidth]{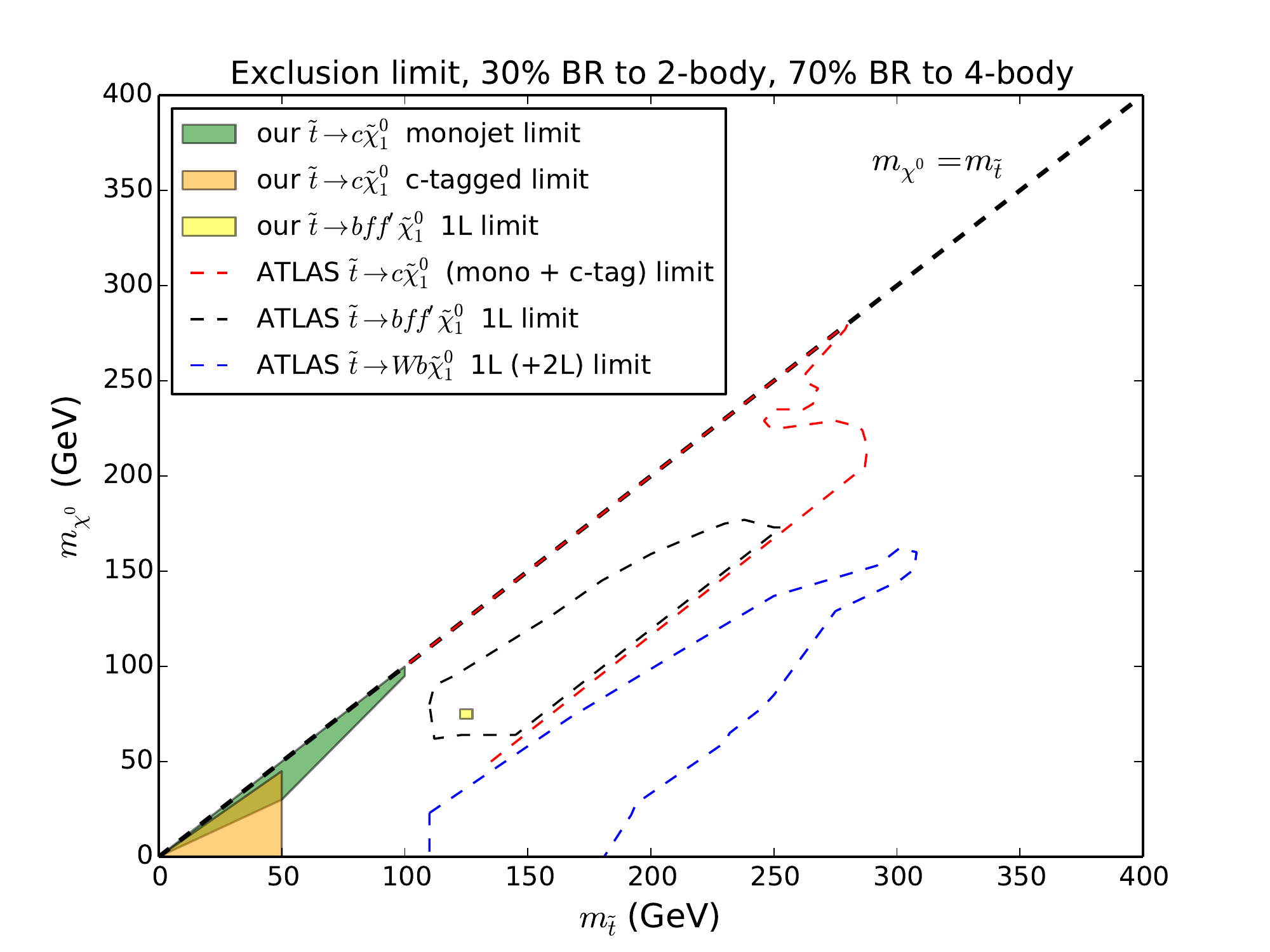}}\\
\makebox[\textwidth][c]{
  \includegraphics[width=0.5\textwidth]{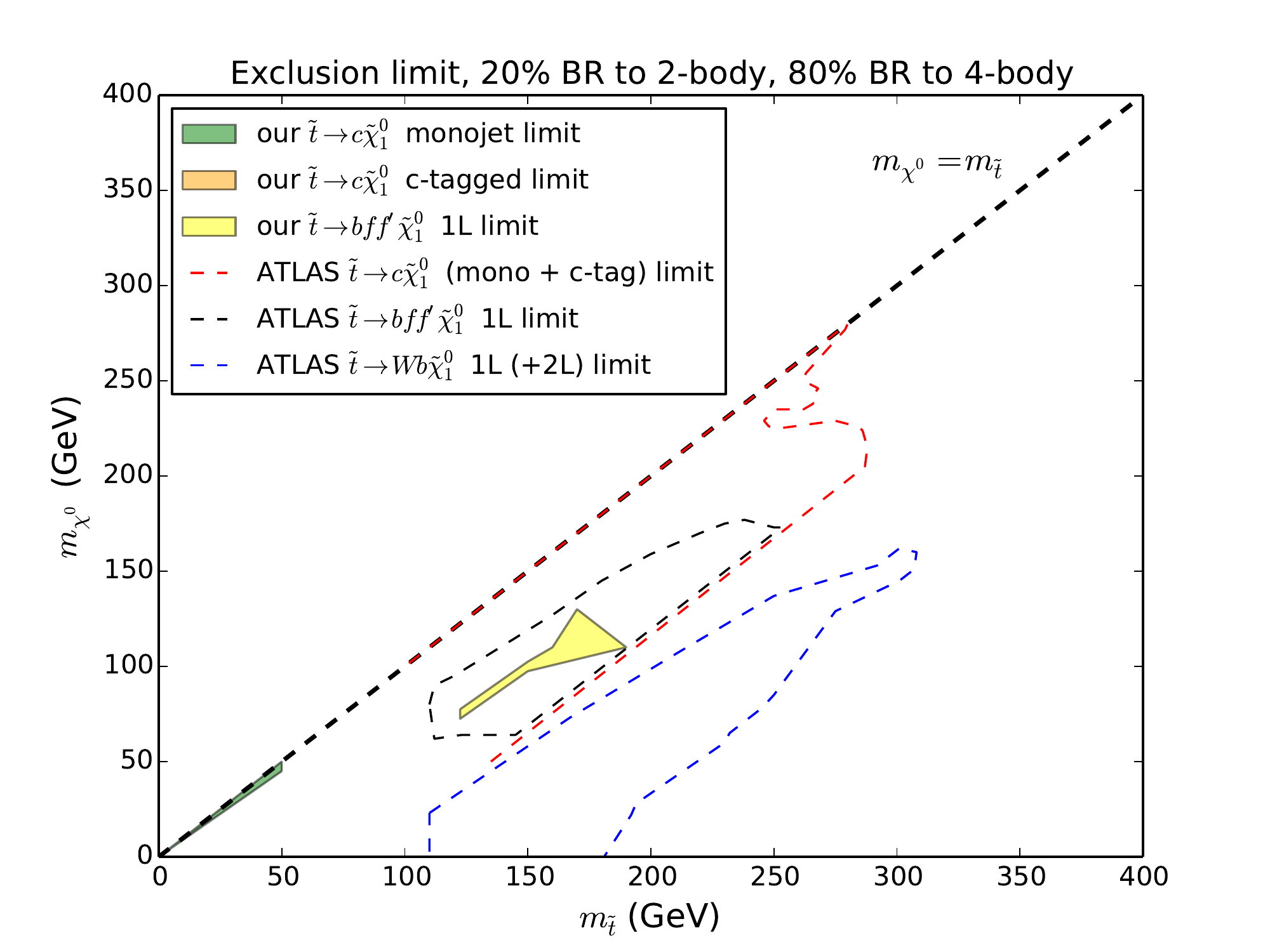}
  \includegraphics[width=0.5\textwidth]{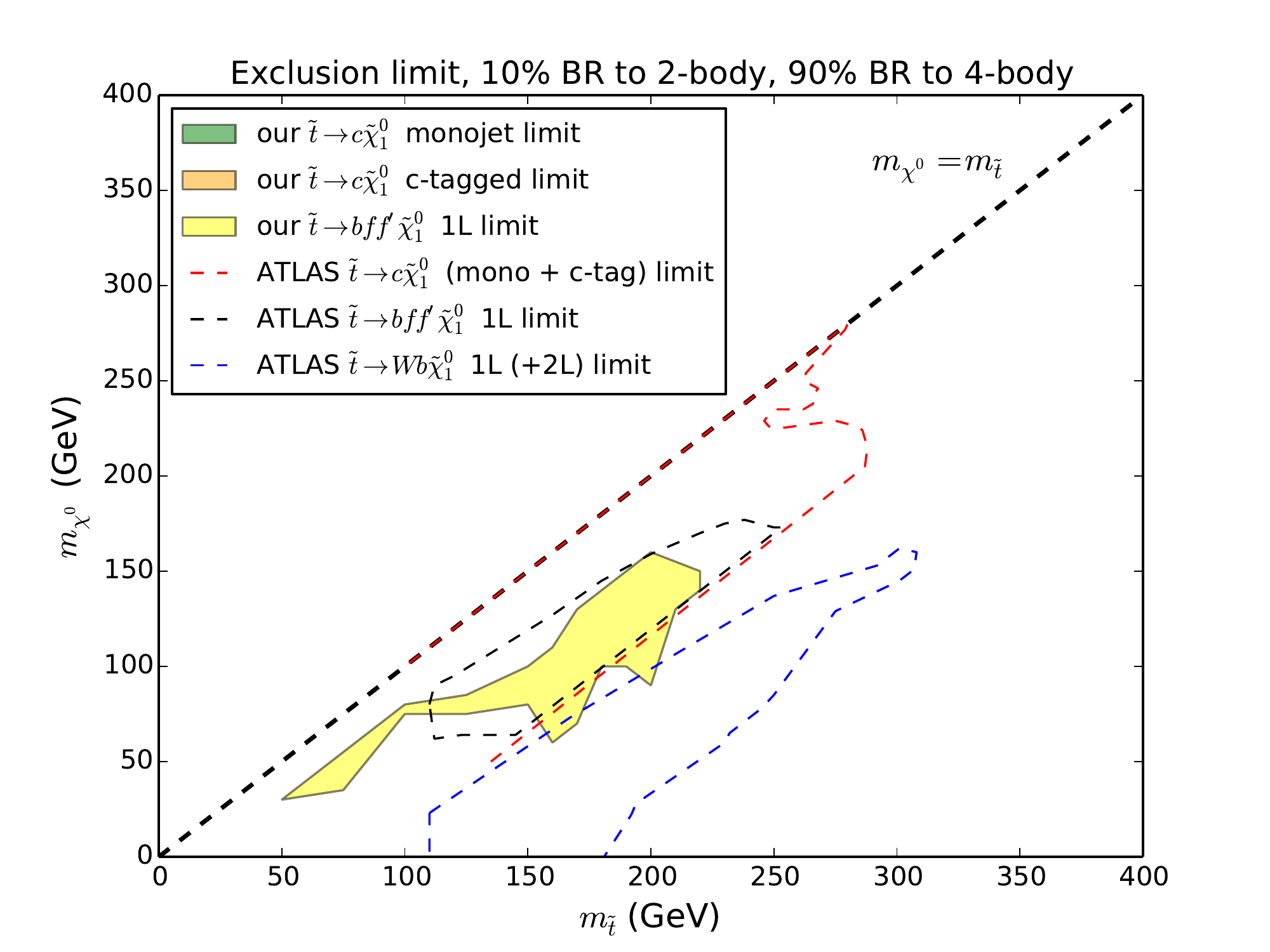}}\\
\makebox[\textwidth][c]{
  \includegraphics[width=0.5\textwidth]{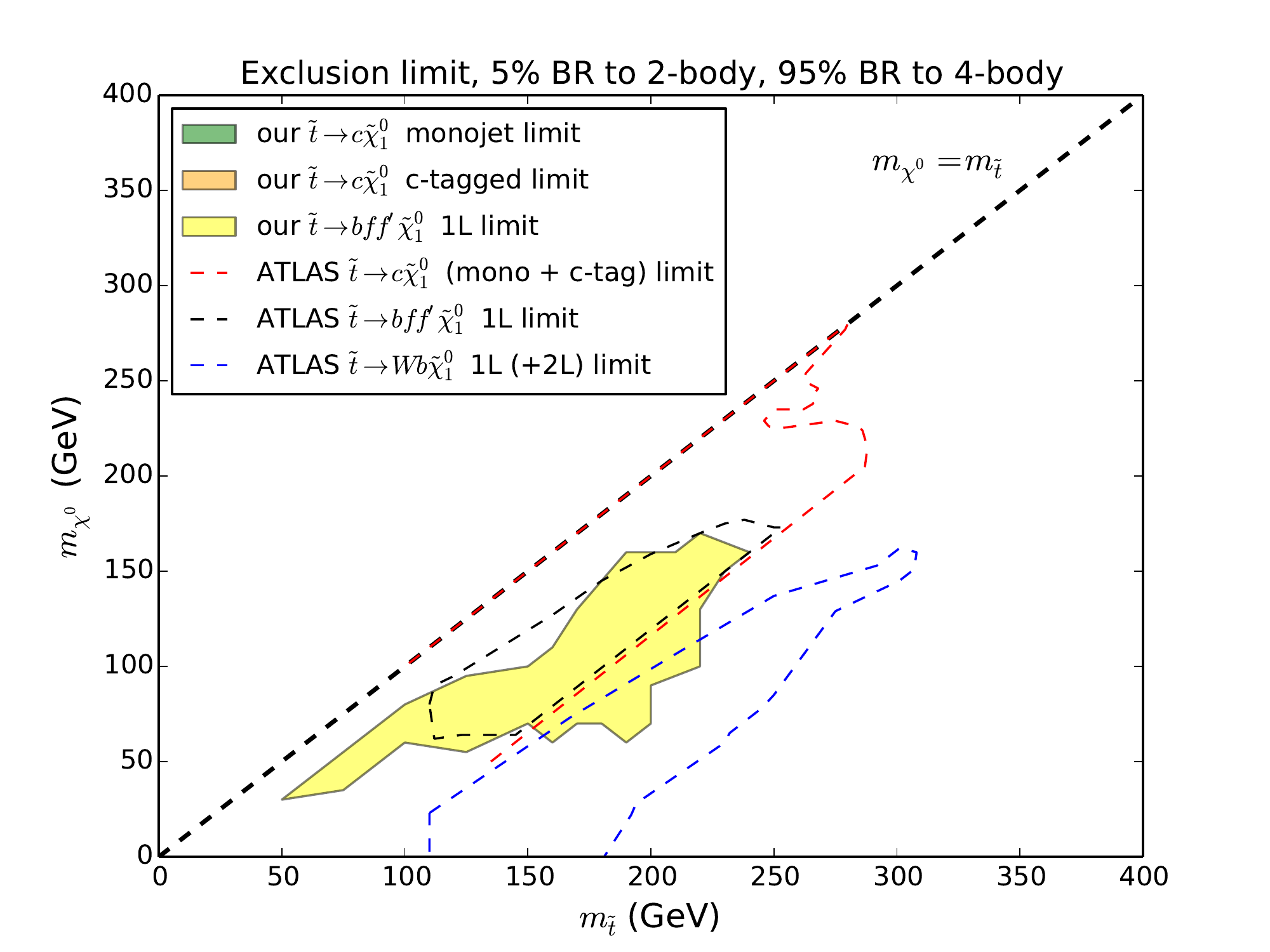}
  \includegraphics[width=0.5\textwidth]{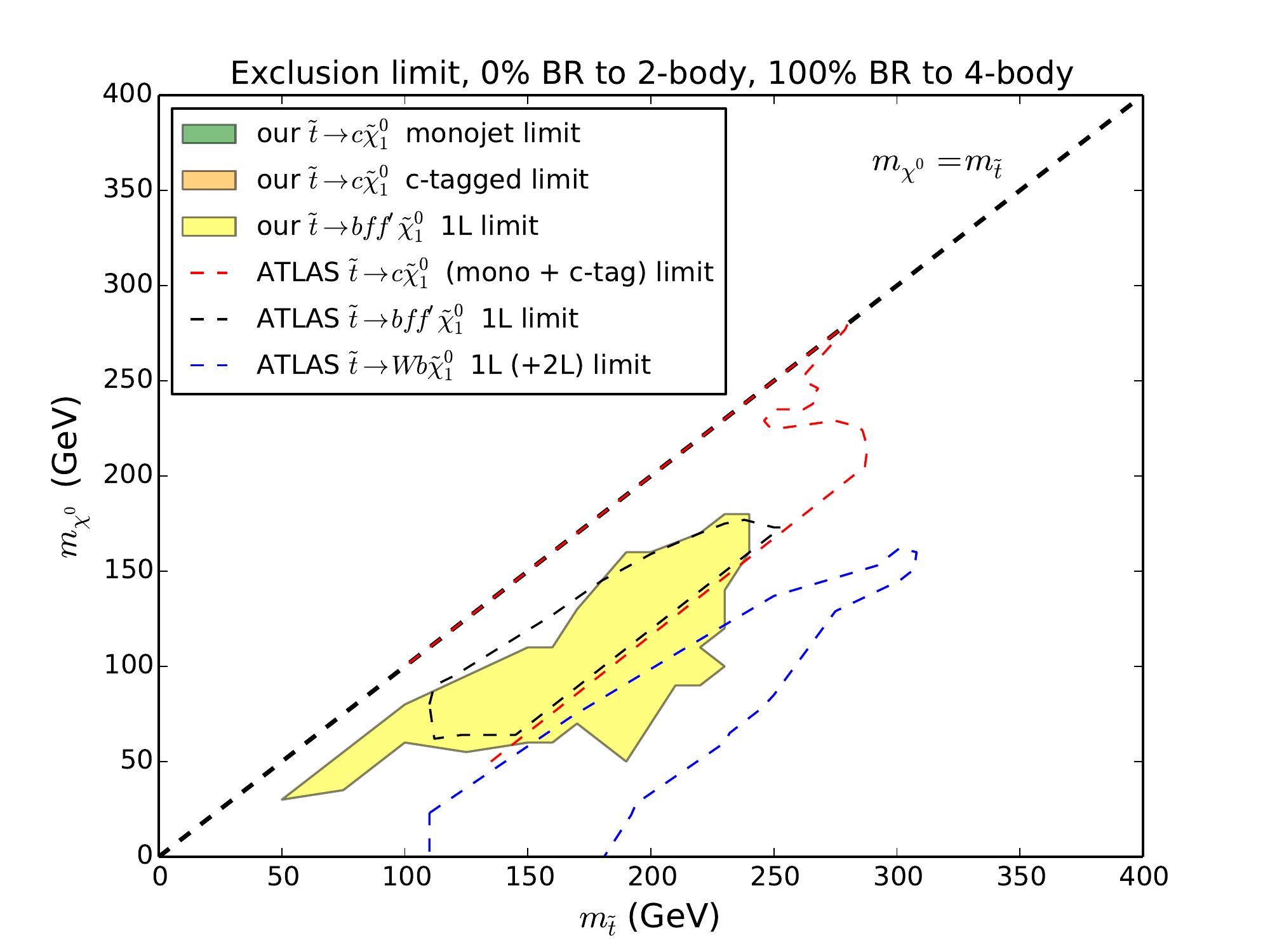}}
\vspace*{-10mm}
\caption{Excluded region assuming certain branching ratios to the 2-body $\tilde{t} \to \tilde{\chi}^0_1 c$ and 4-body $\tilde{t} \to b f f^{\prime} \tilde{\chi}^0_1$ decays. Starting from the top left and working right and down, the BRs to $\tilde{t} \to \tilde{\chi}^0_1 c$ are 100\%, 80\%, 50\%, 30\%, 20\%, 10\%, 5\% and 0\%, with decay being to $\tilde{t} \to b f f^{\prime} \tilde{\chi}^0_1$ otherwise. ATLAS exclusion regions are shown by dashed lines.\label{fig:int_BR}}
\vspace*{-15mm}
\end{figure}

These results are shown in Figure~\ref{fig:int_BR}, where we consider multiple different values of branching ratios (BR).
We can see that when the 2-body branching fraction is between about 10\%--50\%, neither decay is able to exclude our main region of interest where $\Delta m \approx M_W$ between the two ATLAS results. This occurs mainly because the $\tilde{t} \to b f f^{\prime} \tilde{\chi}^0_1$ exclusion region shrinks rapidly as the cross section of this decay channel drops, requiring a BR $> 90\%$ before its 95\% CL extends beyond the $\Delta m = 80$ GeV line. As any combination of branching ratios is possible when $\Delta m \approx M_W$, these plots confirm that we cannot fully exclude these stop masses for every realisation of the MSSM.

Looking more generally at the whole of the LST parameter space, we can ask which values of $m_{\tilde{\chi}^0}$ and $m_{\tilde{t}}$ are excluded in this more realistic scenario of intermediate branching ratios. This is a very important question for the reasons detailed in our introduction. The short answer is that  almost none of the region with $\Delta m \lesssim 100$ GeV is ruled out~\footnote{We did not attempt to reproduce or compensate for altered branching ratios for the ATLAS analyses which focus on the $\Delta m > 100$ GeV region, which is why we do not comment on altered exclusion for this case.}. %
That is, for any point in the $m_{\tilde{t}}$ vs $m_{\tilde{\chi}^0_1}$ plane where $\Delta m \lesssim 100$ GeV, there is a branching ratio such that it is not ruled out by any of the current 8 TeV analyses, and as the BR can be almost any value in the majority of this region (other than $\Delta m \lesssim$ few GeV), we can not conclusively say that the points are excluded. This is shown most obviously in Fig~\ref{fig:int_BR}, when the BR is 30\% to 2-body and 70\% to 4-body. In this case, the probability of both stops decaying via a 2-body decay is only 9\%, so that only points with extremely light stops ($m_{\tilde{t}}< 100$ GeV) and very high cross sections remain excluded, and the BR to 4-body decay is not high enough to rule out any more than a very small region around $m_{\tilde{t}} = 120$ GeV, $m_{\tilde{\chi}^0} = 75$ GeV (in yellow in the Fig~\ref{fig:int_BR}). In fact, these results do not improve upon previous LEP results of $m_{\tilde{t}} > 95$ GeV \cite{Abbiendi:2002mp,Achard:2003ge,Abdallah:2003xe,Heister:2002hp}. 

For our study, we have assumed decays to only 2 channels $(t \to \tilde{\chi}^0_1 c$ or $t \to b W^{(\ast)} \tilde{\chi}^0_1 \to b f f^{\prime} \tilde{\chi}^0_1)$. Going beyond this assumption and also allowing the decay $t \to b \tilde \chi^\pm \to W b \tilde{\chi}^0 \to b f f^{\prime} \tilde{\chi}^0_1$ would likely reduce the excluded region even further.

\section{Conclusions\label{sec:Conclusions}}

In our study we have extended experimental searches to cover specific gaps in the LST parameter space with  $m_{\tilde{t}}<m_{top}$.
We should note that we were able to achieve this as some of the experimental studies have limitations from SUSY signal sample production and analysis, rather than direct limitations from the LHC experiment. 
In particular, we wanted to rule out as much as possible of the 3-dimensional ($m_{\tilde{t}}, m_{\tilde{\chi}^0_1}$ and $\epsilon_{2B}$)
LST parameter space which previously  was not covered completely around   $\Delta m \approx M_W$ even for 
simplified scenarios with $\epsilon_{2B}=1$ or $\epsilon_{2B}=0$ i.e.
cases with 100\% BR for 2-body or 4-body stop decays respectively.

Assuming a 100\% branching fraction of $\tilde{t} \to \tilde{\chi}^0_1 c$
 we agreed well with ATLAS in the region where they had produced results, validating our signal sample production and analysis code for each particular signature.
Using this validated framework, we have
extended the exclusion well beyond the $\Delta m < m_W$ region, successfully ruling out $m_{\tilde{t}} < 240$ GeV irrespective of neutralino mass after combining our results with those of ATLAS, as shown in Fig.\ref{fig:2body_exc}.
This means that if $Br(\tilde{t} \to \tilde{\chi}^0_1 c)=100\%$, the stops are too heavy to mediate EWBG, and therefore we have excluded the light stop EWBG scenario, independently of other considerations such as Higgs measurements.
 
When instead we assume that the stop only decays via $\tilde{t} \to b f f^{\prime} \tilde{\chi}^0_1$, our results again agree well with the ATLAS exclusion limits where they have produced results. We also extend these results, covering an important gap between two ATLAS analyses where $\Delta m \approx M_W$, although a small region where $ m_{\tilde{t}} \approx 120$ GeV with $m_{\tilde{\chi}^0} \approx$ 40 GeV remains unexcluded as one can see from Fig.\ref{fig:4body_exc}. Therefore we have limited the values of $m_{\tilde{t}}$ and reduced the amount of parameter space remaining where light stop EWBG is still viable, although it is not ruled out entirely.

However one should stress that in a more general LST scenario with an intermediate value of $\epsilon_{2B}$ between 0 and 1, the exclusion parameter space can be dramatically different. Having explored this possibility,
we found that if the branching fraction to charm and neutralino is between 10\%--50\% (as illustrated in Fig.~\ref{fig:int_BR}), then our new exclusion limits are much reduced and do not extend beyond $\Delta m > 80$ GeV. In this region, the decay is most likely to be $\tilde{t} \to b f f^{\prime} \tilde{\chi}^0_1$ in a model with no flavour violation (beyond the SM), but more generally any value of BR is possible, and so the most general exclusions limits are much weaker. 

More generally, we have shown that whilst current experimental analyses using simplified models with only one decay channel look like they've excluded the majority of the LST scenario parameter space, in a more general and realistic scenario, allowing just two decay channels dramatically reduces the region which is definitively excluded. In fact in this case, the limits on stop masses is reduced to $m_{\tilde{t}} > 95$ GeV from LEP \cite{Abbiendi:2002mp,Achard:2003ge,Abdallah:2003xe,Heister:2002hp}. This means that given a realistic model, the LST scenario is far from excluded, which has important implications for naturalness as well as allowing stop masses light enough to facilitate electroweak baryogenesis.

The exclusion of the LST parameter space could be further improved
in this more general scenario with a mixture of 2-body and 4-body decays by 
doing a fuller combination/optimisation of the respective signatures.
This study goes beyond the subject of the present paper.

A further complication which we have ignored, but needs to be considered in a completely generic scenario
is allowing for light charginos entering the decay chains. When the LSP neutralino is Higgsino like, the chargino mass is close to the LSP mass, so a chargino decay could appear in the stop decay chain, altering the kinematics and affecting the
LST exclusion results. In this study we have made the assumption that this chargino is above the stop mass, so can be ignored, however
had it been included, it is likely that the stop exclusion limits in the general case would 
be weakened further.

In summary we have successfully extended the ATLAS stop exclusion bounds.
If we assume a $Br(\tilde{t} \to \tilde{\chi}^0_1 c)=100\%$
then we have excluded $m_{\tilde{t}}<m_{top}$ and therefore are the first to exclude light stop EWBG based solely on stop masses from direct searches.
On the other hand, assuming a $Br(\tilde{t} \to b f f^{\prime} \tilde{\chi}^0_1)=100\%$, a small area of parameter space remains.
This result has an important impact on the Higgs signal at the LHC:
If $m_{\tilde{t}}>m_{top}$, then the effect from light stops loops
is expected to be {\it below a few $\%$} for all the main Higgs production and decay
observables~\cite{Belyaev:2013rza}. Furthermore, we demonstrate that for stop BRs different from $100\%$, the excluded region of the LST scenario is dramatically
reduced, so for a generic LST scenario, light stop baryogenesis is still a possibility, necessitating further dedicated studies of the LST parameter space.

\section*{Acknowledgements}

We would like to thank Margarete Muhlleitner for useful discussion regarding this work.
This work is supported
by the Science Technology and Facilities Council (STFC) under grant number
ST/L000504/1.
AB acknowledges support by the STFC under grant number  ST/L000296/1
and by Royal Society Leverhulme Trust Senior Research Fellowship LT140094, and MT acknowledges support from an STFC 
STEP award.

\section*{Note Added}
While finalising our paper, the ATLAS collaboration released a note with new results on the stop
search\cite{Aad:2015pfx}. This extends their previous excluded region under the assumption of a 4-body final state (via $\tilde{t}_1 \to b W^{(*)} \tilde{\chi}^0_1$). However our study remains important, having both extended the excluded region even beyond this new ATLAS result, ruling out heavier stops for 4-body decays, as well as significantly extending the exclusion assuming a 2-body final state which was not addressed in this new ATLAS note.
\newpage
\bibliographystyle{JHEP}
\bibliography{bib}

%\bibliography{mono}{}
%\providecommand{\href}[2]{#2}\begingroup\raggedright
%\begin{thebibliography}{10}
%\bibitem{REVTEX41Control}

%\end{thebibliography}\endgroup

\end{document}